\documentclass[preprint]{ptephy_v1}
\usepackage{hyperref}
\usepackage{graphicx} 
\usepackage{dcolumn}
\usepackage{bm}
\usepackage{amssymb}
\usepackage{amsmath,amsthm}
\usepackage{bm}
\usepackage{mathrsfs}

\newcommand{\be}{\begin{equation}}
\newcommand{\ee}{\end{equation}}
\newcommand{\bea}{\begin{eqnarray}}
\newcommand{\eea}{\end{eqnarray}}
\newcommand{\beqn}{\begin{eqnarray}}
\newcommand{\eeqn}{\end{eqnarray}}
\newcommand{\ba}{\begin{array}}
\newcommand{\ea}{\end{array}}

\newcommand{\noi}{\noindent}

\newcommand{\ra}{\rightarrow}

\usepackage{url}

\newcommand{\beq}{\begin{equation}}
\newcommand{\eeq}{\end{equation}}

\begin{document}
\title{Full Electroweak  $\mathcal{O}(\alpha)$ corrections to Higgs boson production processes with the beam polarization at the International Linear Collider\footnote{This paper is based on the Ph.-D thesis of Nhi M.U. Quach in 2021 \cite{Nhi2021} }}

\author{Nhi My Uyen Quach}
\affil{High Energy Accelerator Research Organization (KEK), Tsukuba, Ibaraki 305-0801, Japan, and also The Graduate University for Advanced Studies (SOKENDAI), Hayama, Kanagawa 240-0193, Japan.}

\author{Junpei Fujimoto}
\affil{High Energy Accelerator Research Organization (KEK), Tsukuba, Ibaraki 305-0801, Japan, and also The Graduate University for Advanced Studies (SOKENDAI), Hayama, Kanagawa 240-0193, Japan.\email{junpei.fujimoto@kek.jp}\\}
 
\author{Yoshimasa Kurihara}
\affil{High Energy Accelerator Research Organization (KEK), Tsukuba, Ibaraki 305-0801, Japan,and also The Graduate University for Advanced Studies (SOKENDAI), Hayama, Kanagawa 240-0193, Japan.\email{yoshimasa.kurihara@kek.jp}\\}

\begin{abstract}
We present the full ${{\cal O}}(\alpha)$
electroweak radiative corrections to 9 Higgs boson
production with the beam polarization at the International Linear Collider(ILC).
The computation is performed with the help of {\tt GRACE-Loop}. 
We compared the full ${{\cal O}}(\alpha)$ electroweak radiative corrections with the factorized initial state radiation(ISR) effects
to estimate the pure weak corrections. This reveals the pure weak corrections are not negligible and should be taken into account for the precision measurements of the Higgs boson property at the ILC experiments.
\end{abstract}
\subjectindex{ILC, Standard Model, Higgs production, recoil mass distriution,Full Electroweak  $\mathcal{O}(\alpha)$ corrections}

\subjectindex{~}
\maketitle
\section{Introduction}
The ILC is a future linear electron-positron collider, operating at 250 to 500 GeV center-of-mass energies with high luminosity which can be extended to 1 TeV in the upgrade stage. It is, based on 1.3 GHz superconducting radio-frequency (SCRF) accelerating technology \cite{Behnke:2013xla},\cite{Fujii:2013lba}. The ILC is proposed to be constructed in the Kitakami Mountains in Tohoku area, Japan. It is an international project running for more than 20 years collaborated by more than 300 institutes, universities, and laboratories. \\

The major physics aim of the ILC is to determine the future direction of particle physics via the precise measurements of the couplings of the Higgs boson with other elementary particles. A wide variety of elementary particles and nuclear physics can be studied. 

On July 4 in 2012, the Higgs boson discovery was announced by physicists from two experiments (the CMS group and the ATLAS group) of the LHC at CERN. A new particle with a mass of approximately 125 GeV and several other properties of the Higgs boson, as predicted by the SM, were observed  in Refs.\cite{Aad:2012tfa} and \cite{Chatrchyan:2012ufa}. The Higgs boson discovery made the SM of particle physics complete. \\

The prediction of additional Higgs bosons is one
of the prominent features of possible physics of beyond the standard model, BSM, which leads to an extended Higgs-boson sector. Searching for higher-mass Higgs bosons is also very interesting. However, the hypothesis that the Higgs boson is heavy and approaches the theoretical upper bound has been validated by the expected exclusion regions with 300 $\text{fb}^{-1}$ and 3 $\text{ab}^{-1}$ LHC data. Therefore, the ILC is an exemplary collider to study the 125 GeV Higgs boson. 
\noindent
With this Higgs particle, the first course at the ILC would be at threshold around $\sqrt{s}= 250$ GeV,
which displays the clear peak cross section for the process $e^{+}e^{-} \rightarrow ZH$. At this energy, the precise measurements of Higgs recoil mass for the Higgs-strahlung process $e^{+}e^{-} \rightarrow ZH$ with subsequent $Z \rightarrow l^+l^- (l= e, \mu)$ decay is one of the most important measurements. This measurement allows a model-independent absolute measurement of the $g_{HZZ}$ coupling.
 In this setting, it is possible to measure the proportion for all of Higgs boson decays including common final states and invisible decays with great accuracy. \\

\noindent
It is not necessary to observe the Higgs decay because corresponding hidden decay is observable. Nevertheless, the $e^{+}e^{-} \rightarrow ZH$ process can be used to measure diverse branching ratios for various Higgs decay processes. the $Z \rightarrow q \bar{q}$ and $Z \rightarrow \nu \bar{\nu}$ processes were included  in this analysis to increase the statistics. The Higgs boson can also decay into a pair of W-bosons. However, the measurement of the WW-fusion process at $\sqrt{s}= 250$ GeV is quite difficult. Conversely, the W-pair production process becomes active at $\sqrt{s}= 500$ GeV, and hence the energy setting of the ILC is expected to be changed, accordingly \cite{Fujii:2013lba}.\\
\noindent

 In this paper, we have studied 9 processes of $e^+e^- \rightarrow f\bar{f} H$ with the polarized electron and positron beams of center-of-mass energy at 250 GeV .i.e. the production of (1) muon pair and Higgs boson, (2) electron-positron pair and Higgs boson, (3) tau-pair and Higgs boson, (4) muon-neutrino pair and Higgs boson, (5) electron-neutrino pair and Higgs boson, (6) up-quark pair and Higgs boson, (7) down-quark pair and the Higgs
boson, (8) charm-quark pair and Higgs boson, and  (9) bottom-quark
pair and the Higgs boson (note that we did not consider the production of top quark pair, strange quark pair, and tau-neutrino pair). We calculated the $\mathcal{O}(\alpha)$ corrections using the on-shell renormalization scheme to each of these 9 reactions as well as the one-photon emission process.

In this work, we have used the {\tt GRACE-Loop} system to calculate the amplitudes automatically. We have developed two sets of Fortran codes. 

The first set was used to calculate the $O({\alpha})$ corrections corresponding to the one-loop diagrams and the one-photon emission processes; in this case, all the fermion masses were kept non-zero, except the neutrino masses, and the arbitrary longitudinal polarization of the input electron and positron beams were available. 
The second set was used to treat the effects of initial-state radiation (ISR). The effects of one-loop weak corrections were determined by comparing the $O({\alpha})$ corrections with the ISR effects.\\

We introduce "NOLFS (NO Light-Fermion Scalar coupling) approximation", where the couplings between the light-fermions (except for the bottom- and top-quarks) and scalar particles are neglected, but the masses of all fermions are retained. We confirmed the one-loop amplitudes generated using the NOLFS approximation to be consistent with those obtained from the Monte-Carlo integration of the full Lagrangian. Therefore, in this work, we present and discuss the results obtained using the NOLFS approximation. \\

The structure of this paper is as follows; after the introduction, we will explain the calculation framework of {\tt GRACE-Loop} in Section II.
In Section III, we will define the beam polarization of the ILC beams. The recoil mass analysis is introduced in Section IV. In Section V, we will explain the radiator method of the initial state radiation. Then we will present the numerical results of the leptonic processes in Section VI and those of quark processes in Section VII. Then we will make a discussion  in Section VIII. Section IX is devoted to the conclusion of this paper.

\section{{\tt GRACE-Loop} and the calculation of $e^{+}e^{-}\rightarrow f\bar{f}H$ }

The {\tt GRACE-Loop} system is a programming system for calculating full one-loop electroweak cross sections and the tree-level automatically, with beam polarizations based on the SM and Minimal Supersymmetric Standard Model (MSSM) of high energy physics. This program was created and developed by the Minami-Tateya group at the High Energy Accelerator Research Organization, KEK \cite{minamilink}. It uses the symbolic manipulation system {\tt FORM} \cite{Vermaseren} to calculate all Dirac and tensor algebra in $n$ dimensions.

In the {\tt GRACE} system \cite{minamilink}, at the tree level, the beam polarization effects are taken into account from the beginning to produce the helicity amplitudes, which are evaluated by a numerical way. Nevertheless, at the full one-loop electroweak corrections with the  {\tt GRACE-Loop} system, it adopts the trace techniques to get the interference between the tree amplitudes and one-loop ones using the symbolical manipulation with the  {\tt FORM} system. Then, the polarization effects of the beams are included with the insertion of the projection operators, $\frac{1\pm\gamma_5}{2}$ into amplitudes. It makes the program size become much larger.

It has been successfully tested for a variety of one-loop $2\ra 2$ electroweak processes \cite{BELANGER2006117}. 
It also provided the first results on the full one-loop radiative corrections to $e^+e^- \ra \nu \bar{\nu} H$ \cite{eennhletter} which has been confirmed by an independent calculation \cite{Denner:2003iy}. 

The {\tt GRACE-Loop} system primarily focuses on evaluating one-loop corrections to the SM processes during electron-positron collisions. In addition, it can calculate one-loop corrections to the MSSM \cite{Junpei:2007qp}.

It should be noted the FULL model contains all of the couplings of the scalar particles such as the Higgs boson or pseudo-Goldstone scalar bosons and fermions in the SM are included. 

 In this paper, the bottom- and top-quark are considered to be heavy fermions and other fermions are recognized to be light ones.

When the two cases of the FULL model and NOLFS approximation are compared, the number of tree diagrams and one-loop Feynman diagrams in the former is much larger than in the latter; thus, the integration over phase space is not practical in the former, as shown in Table \ref{tab:4.1}.\\

\begin{table}[h]
	\begin{center}
		\begin{tabular}{|c|r r r|r r r|}
			\hline
			 &
			\multicolumn{3}{c|}{the NOLFS approximation} &
			\multicolumn{3}{c|}{the FULL model}  \\
			\hline 
			Graph information & one-loop  & Tree  & 5-point & one-loop  & Tree  & 5-point\\
			\hline
			$\mu^+\mu^-H$                 & 208 & 1 & 10 & 2235 & 21 &170\\
			$e^+e^-H$                     & 416 & 2 & 20 & 4470 & 42 &340\\
			$\tau^+\tau^-H$               & 208 & 1 & 28 & 2235 & 21 &188\\
			$\nu_{\mu} \bar{\nu}_{\mu} H$ & 122 & 1 & 6  &  604 &  4 & 36\\
			$\nu_{e} \bar{\nu}_{e} H$     & 219 & 2 & 15 & 1350 & 12 &113\\
			$u \bar{u} H$                 & 209 & 1 & 10 & 2327 & 21 &174\\
			$d \bar{d} H$                 & 209 & 1 & 10 & 2327 & 21 &174\\
			$c \bar{c} H$                 & 209 & 1 & 10 & 2327 & 21 &174\\
			$b \bar{b} H$                 & 651 & 6 & 29 & 2327 & 21 &193\\
			\hline
		\end{tabular}
		\caption{The numbers of Feynman diagrams for the NOLFS approximation and the FULL model of 9 processes}
		\label{tab:4.1}
	\end{center}
\end{table}

From Table \ref{tab:4.1}, we observe that the number of one-loop Feynman diagrams of the $\mu^+\mu^-H$ and $\tau^+\tau^-H$ processes have the same, with 208 and 2235 diagrams with the NOLFS approximation and the FULL model, respectively. Similarly, the number of one-loop Feynman diagrams of $u \bar{u} H$, $d \bar{d} H$ and $c \bar{c} H$ processes have the same, with 209 and 2327 diagrams according to one-loop level with the NOLFS approximation and the FULL model, respectively. However, the numbers of Feynman diagrams one-loop of $b \bar{b} H$ have 651 with the NOLFS approximation and 2327 with the FULL model. \\

For all electroweak processes we
adopt the on-shell renormalisation scheme according
to \cite{kyotorc} and \cite{BELANGER2006117}. The one-loop scalar integrals with the width are evaluated using the {\tt LoopTools} packages \cite{LoopTools}, on the other hand, the detail of the reduction on the tensor integrals and the 5-point functions based on the Feynman parameters are
described in \cite{BELANGER2006117} .

The produced matrix elements to the electroweak corrections by the {\tt GRACE-Loop},  are checked by
performing three kinds of tests at a random point in phase
space. For these tests to be passed one works in quadruple
precision. 

We first have  to check the ultraviolet finiteness
of the results. This test applies to the whole set  of the virtual
one-loop diagrams. In order to conduct this test we regularise any
infrared divergence by giving the photon a fictitious mass (we set
this at $\lambda=10^{-15}$GeV). In the intermediate step of the
symbolic calculation dealing with loop integrals (in
$n$-dimension), we extract the regulator constant
$C_{UV}=1/\varepsilon -\gamma_E+\log 4\pi$, $n=4-2 \varepsilon$
and treat this as a parameter. The ultraviolet finiteness test is
performed by varying the dimensional regularisation parameter
$C_{UV}$. This parameter could then be set to $0$ in further
computation.

By varying $C_{UV}$ from 0 to 100 for the $e^{+}e^{-}\rightarrow \mu^+\mu^-H$ process of the FULL model with $k_c=10^{-1}$ GeV, $\lambda = 10^{-17}$  GeV  
and $\Gamma_Z =0$ GeV at $\sqrt{s}=250$ GeV, we obtained $C_{UV}$ independence with an agreement up tp 33 digits as shown in Table \ref{tab:4.2}.
\begin{table}[h] 
	\begin{center}
		\begin{tabular}{c r}
			
			\hline 
			$C_{UV}$ & Evaluation of the 1 loop amplitude\\
			\hline
			0         &  0.127132019079521763060824923741166\\
			100       &  0.127132019079521763060824923741166\\
			\hline
		\end{tabular}
		\caption{Changing $C_{UV}$ from 0 to 100 of $e^{+}e^{-}\rightarrow \mu^+\mu^-H$  of the FULL model  at a phase point in the arbitrary unit with $k_c=10^{-1}$ GeV, $\lambda = 10^{-17}$ GeV and  $\Gamma_Z =$ 0 GeV  at $\sqrt{s}=250$ GeV. }\label{tab:4.2}
	\end{center}
\end{table}

As the second  test,  the infrared finiteness was performed by
including both loop and bremsstrahlung contributions and checking
that there is no dependence on the fictitious photon mass
$\lambda$. 

Fictitious photon mass($\lambda$) independence was achieved by varying $\lambda$ from $10^{-17}$ GeV to $10^{-19}$ GeV with $C_{UV} = 100 $ and  $\Gamma_Z =0$ GeV  has up to 14-digit agreement as shown in Table \ref{tab:4.4}.

\begin{table}[h]
	\begin{center}
		\begin{tabular}{ c r}		
			\hline 
			$\lambda$ [GeV]  & Evaluation of the 1 loop amplitude\\
			\hline
			$10^{-17}$           &  0.127132019079521763060824923741166\\
			$10^{-19}$           &  0.127132019079522845528273933268792\\
			\hline
		\end{tabular}
		\caption{$\lambda$ independence of $e^{+}e^{-}\rightarrow \mu^+\mu^-H$ of the FULL model  at a phase space point in the arbitrary unit  by changing from $10^{-17}$ GeV to $10^{-19}$ GeV with $C_{UV} = 0 $ at $\sqrt{s}=250$ GeV and  $\Gamma_Z = $0 GeV. }
		\label{tab:4.4}
	\end{center}
\end{table}

A crucial test concerns the gauge
parameter independence of the results. Gauge parameter
independence of the result is performed through a set of five
gauge fixing parameters. For the latter a generalised non-linear
gauge fixing condition \cite{Boudjema:1995cb},\cite{BELANGER2006117} has been
chosen.

\beqn
\label{fullnonlineargauge} {{\cal L}}_{GF}&=&-\frac{1}{\xi_W}
|(\partial_\mu\;-\;i e \tilde{\alpha} A_\mu\;-\;ig c_W
\tilde{\beta} Z_\mu) W^{\mu +} + \xi_W \frac{g}{2}(v
+\tilde{\delta} H +i \tilde{\kappa} \chi_3)\chi^{+}|^{2} \nonumber \\
& &\;-\frac{1}{2 \xi_Z} (\partial.Z + \xi_Z \frac{g}{ 2 c_W}
(v+\tilde\varepsilon H) \chi_3)^2 \;-\frac{1}{2 \xi_A} (\partial.A
)^2 \;. \eeqn The $\chi$ represent the Goldstone. We take the 't
Hooft-Feynman gauge with $\xi_W=\xi_Z=\xi_A=1$ so that no
``longitudinal" term in the gauge propagators contributes. Not
only this makes the expressions much simpler and avoids
unnecessary large cancelations, but it also avoids the need for
high tensor structures in the loop integrals. The use of the five
parameters, $\tilde{\alpha}, \tilde{\beta}, \tilde{\delta},
\tilde{\kappa}, \tilde\varepsilon $ is not redundant as often
these parameters check complementary sets of diagrams.

Non-linear gauge parameter-independence was confirmed by varying $\tilde{\alpha},\tilde{\beta},\tilde{\delta},\tilde{\kappa},\tilde{\varepsilon}$ from $(0,0,0,0,0)$ to $(10,20,30,40,50)$  with $C_{UV}=100$ at  $\sqrt{s}=250$ GeV  and  $\Gamma_Z =$  0 GeV in order to avoid the violation of the gauge invariance, which resulted in an agreement up to 15 digits as shown in Table \ref{tab:4.3}. 

\begin{table}[h]
	\begin{center}
		\begin{tabular}{c r}	
			\hline 
			$\tilde{\alpha},\tilde{\beta},\tilde{\delta},\tilde{\kappa},\tilde{\varepsilon}$  & Evaluation of the 1 loop amplitude  \\
			\hline
			0,0,0,0,0                 &  0.127132019079521763060824923741166\\
			10,20,30,40,50       &  0.127132019079522096127732311288128\\
			\hline
		\end{tabular}
		\caption{Non-linear gauge parameters independence of the FULL model by changing $\tilde{\alpha},\tilde{\beta},\tilde{\delta},\tilde{\kappa},\tilde{\varepsilon}$ at a phase point in the arbitrary unit from (0,0,0,0,0) to (10,20,30,40,50) of $e^{+}e^{-}\rightarrow \mu^+\mu^-H$ with $k_c=10^{-1}$ GeV,  $\lambda = 10^{-17}$ GeV at  $\sqrt{s}=250$ GeV and  $\Gamma_Z =$  0 GeV.}
		\label{tab:4.3}
	\end{center}
\end{table}

We also will show the comparison between the FULL model and the NOLFS approximation at a phase space point in Table \ref{tab:4.5}. 
 
 \begin{table}[h]
	\begin{center}
            \begin{tabular}{c r}		
			\hline 
			FULL model v.s.NOLFS approximation & Evaluation of the 1 loop amplitude \\
			\hline 
			FULL model            &    0.127132019079521763060824923741166 \\
			NOLFS approximation           &  0.127180115426583029147877823561430 \\
			\hline
		\end{tabular}
		\caption{Comparison of the 1 loop amplitudes at a phase point between the FULL model and the NOLFS approximation in the arbitrary unit for the  $e^{+}e^{-}\rightarrow \mu^+\mu^-H$ with $k_c=10^{-1}$ GeV  at $\sqrt{s}=250$ GeV and  $\Gamma_Z =$  0 GeV in order to avoid the violation of the gauge invariance.}
		\label{tab:4.5}
	\end{center}
\end{table}

An additional stability test on the NOLFS approximation concerns the
bremsstrahlung part. It relates to the independence in the
parameter $k_c$ which is a soft photon cut parameter that
separates soft photon radiation  and the hard photon performed  by
the Monte-Carlo integration package of the BASES \cite{KAWABATA1995309}. 
Here, we have implemented the finite fixed width  in all Z-boson propagators as $P_Z(q)=[q^2-m_Z^2+im_Z\Gamma_Z]^{(-1)}$ with $\Gamma_Z=2.49$ GeV \cite{PhysRevD.98.030001} in order to avoid hitting  the pole of the Z-boson.propagators  in the amplitudes.

In the BASES operation, to integrate one-loop amplitude,  we set 10 iteration steps for the grid optimization
and 200 iteration steps for the accumulation with 40,000 sampling points. it takes 16 hours for $\mu^+\mu^-H$ and other
processes; almost 103 hours for $b \bar{b} H$, with 16 Xeon 3.20 GHz CPU
with the size of memory 128 GB. The typical integration errors are $0.06\%$ for
the each step of the loop calculations.\\

\begin{table}[h]
	\begin{center}
		\begin{tabular}{|c||r|r|r|}
			\hline
			    &
			\multicolumn{3}{c|}{$\delta_{Total} (\%)$} \\
			\hline 
			Processes & $k_c=10^{-1}$ GeV  & $k_c=10^{-3}$ GeV & $k_c=10^{-5}$ GeV\\
			\hline
			$\mu^+\mu^-H$                 & $-4.14(08)$    &	$-4.14(08)$  &	$-4.23(08)$\\ 
			$e^+e^-H$                     & $-4.45(09)$           &	$-4.53(09)$  &	$-4.44(09)$\\
			$\tau^+\tau^-H$               & $-4.54(08)$           &  $-4.62(08)$  &   $-4.59(08)$\\
			$\nu_{\mu} \bar{\nu}_{\mu} H$ & $-4.37(13)$           &  $-4.40(14)$  &   $-4.45(16)$\\
			$\nu_{e} \bar{\nu}_{e} H$     & $-3.56(12)$           &  $-3.55(12)$  &   $-3.55(12)$\\
			$u \bar{u} H$                 & $-6.75(10)$           &	$-6.75(10)$  &   $-6.70(10)$\\
			$d \bar{d} H$                 & $-4.97(10)$           &	$-5.06(10)$  &   $-5.12(10)$\\
			$c \bar{c} H$                 & $-6.41(10) $          &	$-6.39(10)$  &   $-6.46(10)$\\
			$b \bar{b} H$                 & $-6.50(11) $          &	$-6.49(11)$  &    $-6.44(11)$\\
			\hline
		\end{tabular}
		\caption{$\delta_{Total}$ in \%  with various $k_c$ values , $k_c$=$10^{-1} $ GeV, $k_c$=$10^{-3} $ GeV, and $k_c$=$10^{-5} $ GeV of all 9 processes without beam polarizations at $\sqrt{s}=250$ GeV.}
		\label{tab:4.6}	
			\end{center}
\end{table}
\noindent
From this table, we claim that the accuracy of our $\mathcal{O}(\alpha)$  corrections is approximately 0.1\% for all 9 processes.

\vspace{1 cm}
As final checks, we will show the comparison between the unpolarized cross sections of $e^{+}e^{-}\rightarrow \nu \bar{\nu}H$ between the current results and those of German group \cite{Denner:2003iy}
 at center-of-mass energy of $\sqrt{s}=500$ GeV because in those days, the Higgs boson was not yet confirmed.

\begin{table}[h]
        \begin{center}
                \begin{tabular}{|c|r|r|r|r|r|r|}
                        \hline
                        $m_H$ (GeV) & $m_W$ (GeV)  & $\sigma_{Tree}$ (pb)  & $\sigma_{\mathcal{O}_\alpha}$ (pb) & $\delta_{Total} \%$ &  \\
                        \hline
                        150 & 80.3767 & 6.1072(9)$\times$$10^{-2}$ & 6.075(6)$\times$$10^{-2}$   & $-0.5$ & Current\\
                        &         & 6.1076(5)$\times$$10^{-2}$  & 6.080(2)$\times$$10^{-2}$    & -0.4 & Ref.\cite{Denner:2003iy}\\
                        200 & 80.3571 & 3.7302(5)$\times$$10^{-2}$  & 3.705(4)$\times$$10^{-2}$    & $-0.7$ & Current\\
                            &         & 3.7293(3)$\times$$10^{-2}$  & 3.709(2)$\times$$10^{-2}$    &$-0.6$ & Ref.\cite{Denner:2003iy}\\
                        250 & 80.3411 & 2.110(2)$\times$$10^{-2}$   & 2.059(2)$\times$$10^{-2}$    & $-2.4$ & Current\\
                        &         & 2.1134(1)$\times$$10^{-2}$  & 2.060(1)$\times$$10^{-2}$    & $-2.5$ & Ref.\cite{Denner:2003iy}\\
                        300 & 80.3275 & 1.0744(7)$\times$$10^{-2}$  & 1.0258(7)$\times$$10^{-2}$ & $-4.5$ & Current\\
                        &         & 1.07552(7)$\times$$10^{-2}$ & 1.0282(4)$\times$$10^{-2}$   & $-4.4$ & Ref.\cite{Denner:2003iy}\\
                        350 & 80.3158 & 4.6077(4)$\times$$10^{-3}$  & 4.172(2)$\times$$10^{-3}$    & $-9.5$ & Current\\
                        &         & 4.6077(2)$\times$$10^{-3
                        }$  & 4.181(1)$\times$$10^{-3}$    & $-9.3$ & Ref.\cite{Denner:2003iy}\\
                        \hline
                \end{tabular}
                \caption{Comparison between the unpolarized cross sections of $e^{+}e^{-}\rightarrow \nu \bar{\nu}H$ between the current results and those of German group \cite{Denner:2003iy} at $\sqrt{s}=500$ GeV.}
                \label{tab:4.12}
        \end{center}
\end{table}

 \newpage
\section{Beam polarization}

Here we would like to define the beam polarization and the realistic polarization status of the proposed ILC.
\hspace{0.5 cm}
The left-handed polarization degree of the electron beam is defined as,
\begin{equation}
p_e =  (N_{e_R} - N_{e_L} )/(N_{e_L} + N_{e_R} ),
\end{equation}
\noindent
where $N_{e_R} $ and $N_{e_L} $ are the number of the right-handed and left-handed electrons in the beam, respectively ,\\
\noindent
and 
\begin{equation}
p_p = (N_{p_R} - N_{p_L} )/(N_{p_R} + N_{p_L} ),
\end{equation}
 $N_{p_R}$ and $N_{p_L}$ are the number of the right-handed and left-handed positrons in the beam, respectively.\\
\noindent
When a normalization $N_L + N_R = 1$ is used, the normalized number of left-handed and right-handed electrons can
be obtained as $N_L = \frac{1 - p_e }{2}$ and $N_R = \frac{1 + p_e }{2}$ , respectively. \\

Therefore, the cross sections for a given combination of the electron and positron beam polarization can be written as:

\begin{equation}
\begin{split}
\sigma(p_e,p_p) = &\frac{1}{4}\{(1-p_e)(1+p_p)\sigma_{LR}  + (1+p_e)(1-p_p)\sigma_{RL}\\
                           &+ (1-p_e)(1-p_p)\sigma_{LL}+ (1+p_e)(1+p_p)\sigma_{RR}\}
\end{split}
\end{equation}
\noindent
where $\sigma_{LR}$ stands for the cross section with the $100 \% $ left-handed polarized electron ($p_e = -1 $) and the $100 \%$ right-handed polarized positron ($p_p = +1 $) 
beams. The cross sections $\sigma_{RL}$, $\sigma_{LL}$ and $\sigma_{RR}$ are defined analogously.\\

\noindent
The ILC proposed polarizations of the design values are
\begin{equation}
p_e = \frac{0.1-0.9}{1}= -0.8,
\end{equation} 
\noindent
and
\begin{equation}
p_p = \frac{0.65-0.35}{1}=0.3.
\end{equation} 
\noindent	
In the following discussions, we use the following short notations for the various polarization conditions;
\begin{itemize}
\item $p_e = 0, p_p = 0$  ~~~~~~~~~UP: unpolarized,
\item $p_e = +1, p_p = +1$ ~~~~RR: right $e^-$ right $e^+$,
\item $p_e = -1, p_p = -1$ ~~~~LL: left $e^-$ left $e^+$,
\item $p_e = +1, p_p = -1$ ~~~~RL: right $e^-$ left $e^+$,
\item $p_e = -1, p_p = +1$ ~~~~LR: left $e^-$ right $e^+$,
\item $p_e = - 0.8, p_p = 0.3$ ~ ILC: the ILC proposed polarization.
\end{itemize}	
\noindent	

\section{Recoil mass analysis}

We introduce the recoil mass analysis that we conducted to make a model-independent measurement of the coupling between the Higgs and Z bosons, (i.e., $g_{HZZ}$ coupling),
using the recoil mass distribution in $e^+e^-\rightarrow ZH$ with $Z\rightarrow\mu^+\mu^-$ \cite{Keisuke}.\\

In electron-positron collisions at $\sqrt{s}=$250 GeV, the main Higgs production mechanism is the Higgs-strahlung process $e^+e^-\rightarrow HZ$. For $m_H = 125$ GeV, the cross section for the s-channel process is maximal close to  $\sqrt{s}=$ 250GeV\\
\noindent
The total HZ cross section is proportional to the square of the coupling between Higgs and Z bosons, that is, $\sigma(e^+e^-\rightarrow HZ)\propto g^2_{HZZ}$, and the cross sections of the decays to the final state in $H\rightarrow X\bar{X}$ can be expressed as 
\begin{eqnarray*}
\sigma(e^+e^- \rightarrow HZ) \times BR (H \rightarrow X \bar{X}) \propto \frac{ g^2_{HZZ}\times g^2_{HXX}}{\Gamma_H}
\end{eqnarray*}
\noindent
In this study, the cross section of $e^+e^-\rightarrow HZ$ was measured using the recoil mass technique. We considered $\sigma(e^+e^-\rightarrow HZ)$, and the recoil mass can be expressed as 
\begin{equation}
m^2_{rec}= s-2\sqrt{_s}(E_{\mu^+} + E_{\mu^-}) + m^2_{\mu^+ \mu^-},
\end{equation}
\noindent
where $\sqrt{s}$ is the center-of-mass energy, and $E_{\mu^+}$ and $E_{\mu^-}$ are the energies of the two muons and $m^2_{\mu^+ \mu^-}$ is the invariant mass of muon and anti-muon from Z decay.

At $\sqrt{s}=250$ GeV, where the energy of the muons from Z decay approximately scales as $\sqrt{s}$, the width of recoil mass distribution increases significantly with increasing center-of-mass energy. Therefore, the leptonic (in particular, muonic) recoil mass analysis leads to a higher precision on $g_{HZZ}$ for $\sqrt{s}=250 $ GeV, where $\sigma(HZ)$ is the largest and the recoil mass peak is relatively narrow.\\
\noindent
Using this technique, one can determine the absolute branching ratios of Higgs boson decays, including those of the invisible decays.\\
\begin{figure}[h!]
\begin{center}
\includegraphics[height =3.5in, angle =0]{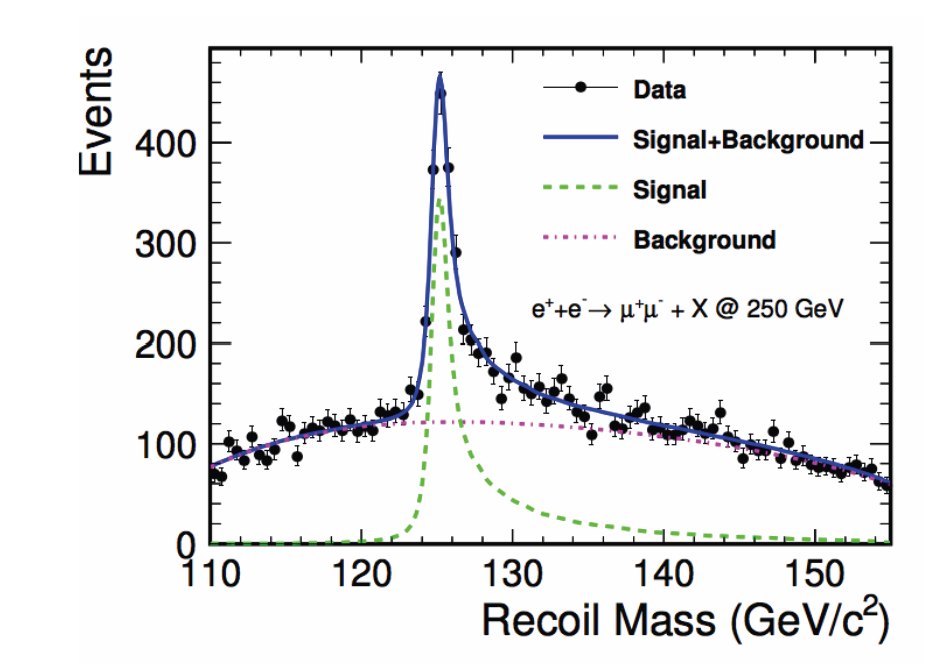}
\caption{Recoil mass distribution of the Higgs-strahlung process $e^{+}e^{-}\rightarrow \mu^+\mu^-H$ ($e^{+}e^{-} \rightarrow ZH$ followed by $Z \rightarrow \mu^+\mu^-$) with 250 $\text{fb}^{-1}$ for $m_h$ = 120 GeV at $\sqrt{s} = 250$ GeV \cite{Keisuke}.}		\label{fig.4}
\end{center}
\end{figure}
\noindent
It should be noted that in this recoil mass distribution, the simple ISR and the detector resolution effects are included, which leads to the appearance of the tail structure in Fig. \ref{fig.4}.

The International Linear Collider (ILC) is a proposed international $e^+e^-$ linear collider with beam energies ranging from 250 GeV to 1 TeV. ILC is supposed to start at the center-of-mass energy of 250 GeV in its initial stage to obtain high statistics measurements of the Higgs boson. Note that this center-of-mass energy was used for the calculations in this paper.

Owing to parity violation in weak interactions, beam polarization effects are essential to resolve new phenomena beyond the standard model (BSM), and hence, studying these effects is necessary. Beam polarization and its importance in studying physics at the $e^+e^-$collider have been discussed in detail over the past decades. A precise measurement to study the properties of the Higgs boson is one of the key targets of the ILC experiments. However, to achieve this, one needs to know the $\mathcal{O}(\alpha)$ corrections to the underlying processes.

\section{Radiator method}

The effect of the initial photon emission can be factorized when the total energy of the emitted photons is sufficiently small compared to the beam energy or for small angle (co-linear) emissions.This approximation is referred as to the "soft-colinear photon approximation(SPA)".
Under SPA, the corrected cross sections with ISR, that is, $\sigma_{ISR}$, can be obtained from the tree cross sections $\sigma_{Tree}$ using a radiator function $H(x,s)$ as follows:
\begin{eqnarray} 
	\sigma_{ISR}&=&\int^1_0 dx~H(x,s)\sigma_{Tree} \left(s(1-x)\right),\label{ISRTree}
\end{eqnarray}
where $s$ is the square of the CM energy and $x$ is the energy fraction of an emitted photon.\\

The total cross section with higher-order QED corrections to ISR can be calculated using the following function  \cite{Fujimoto:2018mfn};
\begin{eqnarray} 
	\sigma_{ISR}&=&\int^{k^2_c/s}_0 dx_1 \int^{1-x_1}_0dx_2~D(x_1,s)D(x_2,s)\sigma_{Tree}\left(sx_1x_2\right).
\end{eqnarray}
wherer the structure function $D(x,s)$, which is corresponding to square root of the radiator function $H(x,s)$,  gives a probability to emit a photon with energy fraction of $x$ at the CM energy square $s$.

In this method, electron and positron can emit different energies, and thus finite boost of the CM system can be treated.
The structure function can be obtained as 
\begin{eqnarray} 
	D(1-x,s)^2&=&H(x,s)=\Delta_2\beta x^{\beta-1}
	-\Delta_1\beta\left(1-\frac{x}{2}\right)\nonumber\\
	&~&+\frac{\beta^2}{8}\left[
	-4(2-x)\log{x}-\frac{1+3(1-x)^2}{x}\log{(1-x)}
	-2x \right],\nonumber\\
	\label{ISRLoop}
\end{eqnarray}
where
\begin{eqnarray*}
	\beta&=&\frac{2\alpha}{\pi}\left(\log{\frac{s}{m_e^2}}-1\right),\\
	\Delta_2=1+\delta_1,&~&\Delta_1=1+\delta_1\\
	\delta_1&=&\frac{\alpha}{\pi}\left(\frac{3}{2} \text{log}\frac{s}{m^2_e} +\frac{\pi^2}{3}-2\right).
\end{eqnarray*}

\noindent
This result is obtained based on perturbative calculations of initial-state photon emission diagrams up to one-loop order.
The Ref.\cite{Fujimoto:2018mfn} shows the formula for the $\mathcal{O}(\alpha^2)$ ISR effects. 
On the other hand, we are interested in separating the $\mathcal{O}(\alpha)$ photonic corrections from the full $\mathcal{O}(\alpha)$  electroweak corrections.
It is known that such a separation can not be realized based on the diagrammatic calculations due to the $SU(2)_L{\times}U(1)$ gauge invariance. 
On the other hand, the exponentiation method has no such problems estimating the initial state photonic effects. 
However, it is exponentiated because we would like to extract $\mathcal{O}(\alpha)$ photonic effects as precisely as possible. 
We neglect the term $\delta_2$ which is just the term of the $\mathcal{O}(\alpha^2)$ ISR formula in the Ref.\cite{Fujimoto:2018mfn}.

\section{Results of the lepton processes with the polarization}

 A canonical set of the input for the electroweak parameters for the numerical evaluations are chosen as follows;

\begin{table}[h]
\begin{center}
		\begin{tabular}{|c||r||c||r||c||r|}	
			\hline
$m_e$     & $0.51099906\times 10^{-3}$GeV\cite{PhysRevD.98.030001} &  $m_u$  &  0.0063 GeV \cite{10.1143/PTP.68.2134} &  $m_Z$  & 91.1876 GeV\cite{PhysRevD.98.030001} \\
$m_\mu$ & $105.6583389\times 10^{-3}$ GeV \cite{PhysRevD.98.030001}&$m_d$ & 0.0063 GeV \cite{10.1143/PTP.68.2134}   &  $m_W$ & 80.366 GeV \cite{10.1143/PTP.68.2134} \\
$m_\tau$  &1.771 GeV\cite{PhysRevD.98.030001}  &$m_c$ & 1.5 GeV \cite{10.1143/PTP.68.2134}& $m_H$ & 125.1 GeV\cite{PhysRevD.98.030001} \\
$m_{{\nu} e}$      &       0 GeV &	$m_s$ & 0.0094 GeV \cite{10.1143/PTP.68.2134}& $\alpha$ &  1/137.035999084\cite{PhysRevD.98.030001} \\
$m_{{\nu} \mu}$&       0 GeV &	$m_b$ & 4.7 GeV \cite{10.1143/PTP.68.2134}& $\sin^2{\theta_W}$ & $1- \frac{m_W^2}{m_Z^2}$\\
$m_{{\nu} \tau}$&       0 GeV &  	$m_t$ & 172.9 GeV \cite{PhysRevD.98.030001} & $\Gamma_Z$ & 2.49 GeV \cite{PhysRevD.98.030001} \\
			\hline
		\end{tabular}
	\caption{Canonical input parameters}\label{ElevelC60}
	\end{center}
\end{table}

Here the light quark masses are taken from the numerical program {\tt MW.f} written by Z.Hioki \cite{10.1143/PTP.68.2134} to reproduce the total hadronic cross sections of $e^+e^-$ collisions in the low energy regions by means of  the dispersion relations, and which also calculates the electroweak one-loop corrected $m_W$ from $\alpha,m_Z, m_{Higgs},m_{e}, m_{\mu}, m_{\tau},$ 
$ m_u,m_d,m_s,m_c$,$m_b$ and $m_{t}$ as inputs parameters to be $m_W=$  80.366 GeV.

\subsection{$e^{+}e^{-}\rightarrow \mu^+\mu^-H$}

\begin{figure}[h!] 
	\begin{center}
		\includegraphics[width=5.0in, height =2.5in, angle =0]{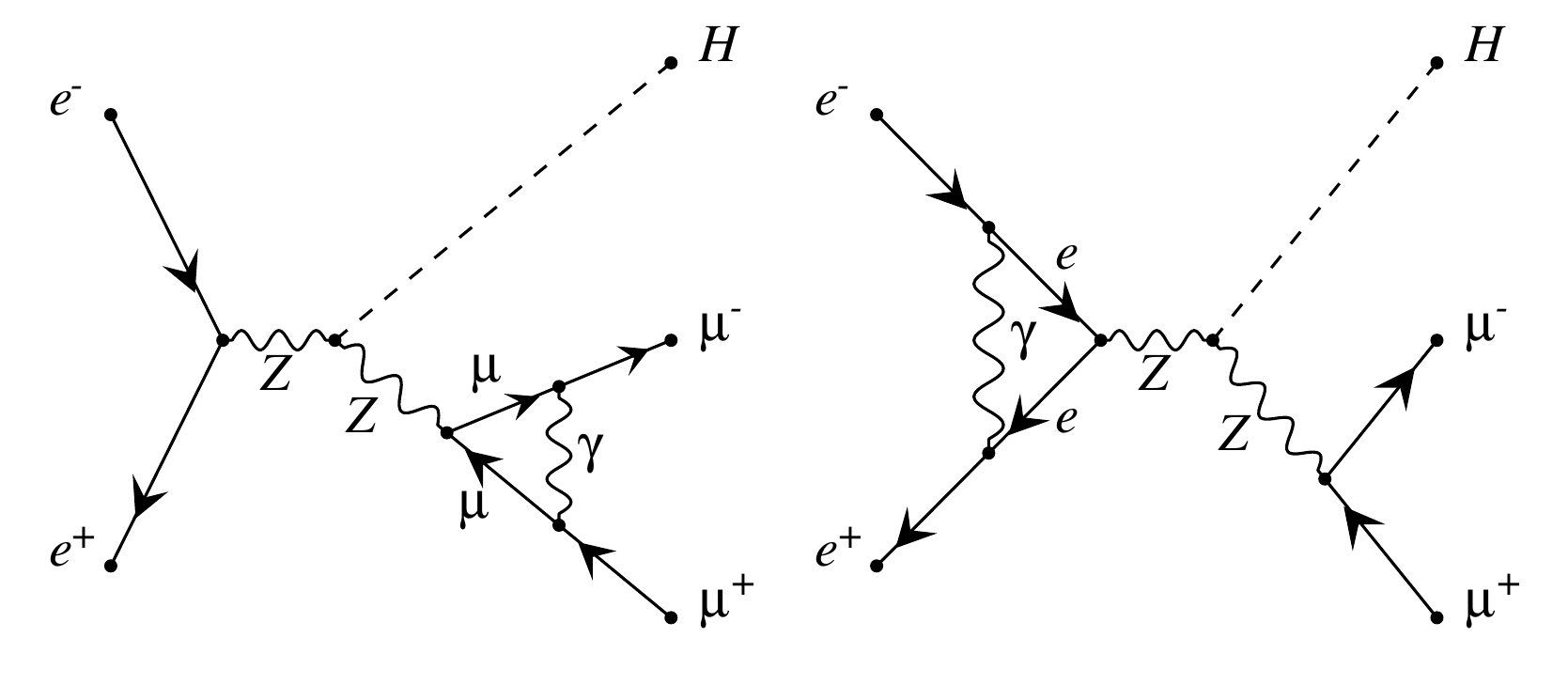}
		\caption{A typical final and initial vertex correction Feynman diagrams of  $e^{+}e^{-}\rightarrow \mu^+\mu^-H$ }
		\label{fig.1}
	\end{center}
\end{figure}

\begin{figure}[h!]
	\begin{center}
		\includegraphics[width=5.0in, height =2.5in, angle =0]{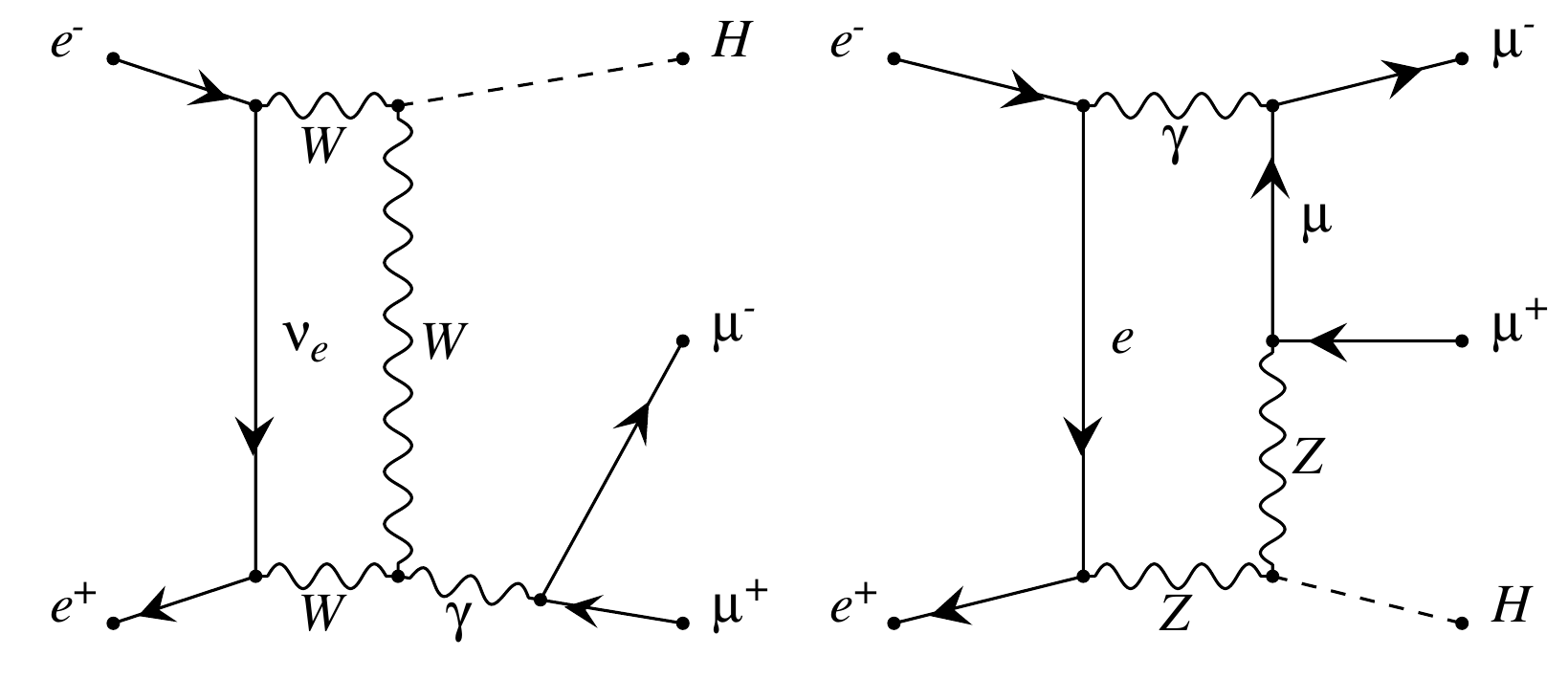}
		\caption{Typical 4-point and 5-point functions of Feynman diagrams of  $e^{+}e^{-}\rightarrow \mu^+\mu^-H$}
		\label{fig.2}
	\end{center}
\end{figure}

\begin{figure}[h!]
	\begin{center}
		\includegraphics[width=2.5in, height =2.5in, angle =0]{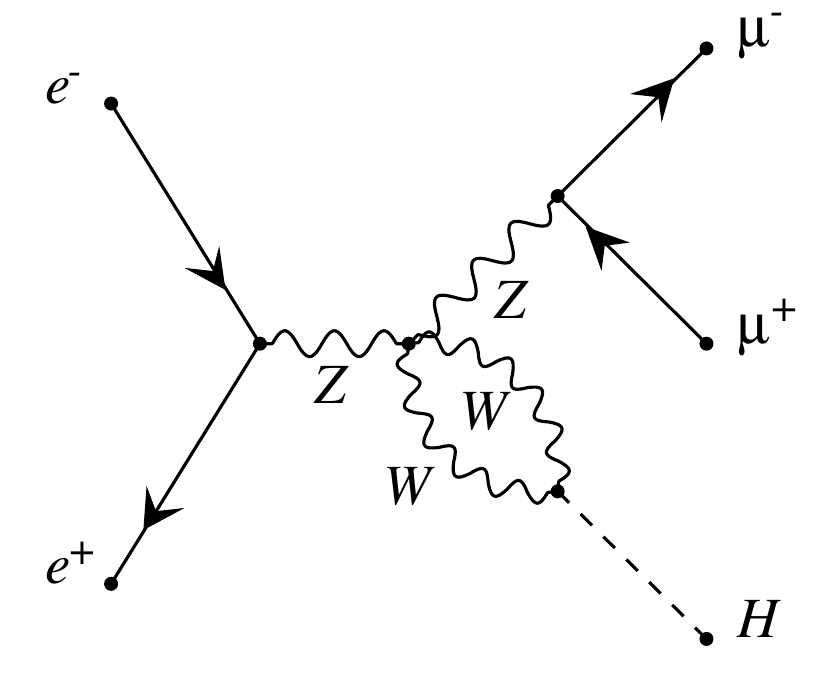}
		\caption{A typical Fish-type Feynman diagram of  $e^{+}e^{-}\rightarrow \mu^+\mu^-H$ }
		\label{fig.3}
	\end{center}
\end{figure}

Now we are going to show five types of typical one-loop Feynman diagrams with typical three-point functions in Fig. \ref{fig.1}, a typical four-point one, a typical five-point one in Fig. \ref{fig.2} and a typical  fish type diagrm in Fig. \ref{fig.3}, respectively.The treatment of these kinds of n-point Feynman diagrams in {\tt GRACE-Loop} is discussed in detail in  \cite{BELANGER2006117,Kiemthesis}.

Table \ref{tab:4.13} shows the total cross sections of the tree level, ones of the ${\mathcal{O}_\alpha}$ corrected with the NOLFS approximation and ones including the ISR effects.
It shows also the total ratios, $\delta_{\mathcal{O}_\alpha}$ with the NOLFS approximation and $\delta_{ISR}$ with the ISR effects for various polarization conditions (i.e., UP, RR, LL, RL, LR, and ILC) defined in  Section $3$. The convergence of phase space integration both of the one-loop effects and also of the ISR effects are very  bad for the case of $LL$ and $RR$ because of the tiny cross section of the tree  in $~10^{-11}$  pb  due to the very  small electron mass effects. Eventhough, for the case of the ILC proposed polarization with $p_e = -0.8$ , $p_p = 0.3$, the convergence of the phase space integration is very good and we get the reliable numbers in $0.1$\% accuracy .\\

\begin{table}[h]
        \begin{center}
                \begin{tabular}{|c||r|r|r|r|r|r|}
                        \hline
                        &      $\sigma_{Tree}$ (pb)    & $\sigma_{\mathcal{O}_\alpha}$ (pb)   &  $\delta_{Total} \%$      & $\sigma_{ISR }$ (pb)    & $\delta_{ISR} \%$ \\
                        \hline
                        UP     & 7.021(2)$\times$$10^{-3}$  &  6.724(1)$\times$$10^{-3}$  & $-4.2$ &6.312(4)$\times$$10^{-3}$  & $-9.9$ \\
                        RR     & 2.16(1)$\times$$10^{-11}$   &  8.4(2)$\times$$10^{-7}$    & 3.91$\times$$10^{6}$  & 9.62(3)$\times$$10^{-13}$  &  $-95.6$ \\
                        LL     & 2.16(1)$\times$$10^{-11}$   &  8.45(2)$\times$$10^{-7}$    & 3.91$\times$$10^{6}$  & 7(2)$\times$$10^{-11}$  &  224.1 \\
                        RL     & 1.108(2)$\times$$10^{-2}$  &    1.194(1)$\times$$10^{-2}$&  7.7 & 1.534(3)$\times$$10^{-3}$ & $-19.3$ \\
                        LR     & 1.709(2)$\times$$10^{-2}$  & 1.497(1)$\times$$10^{-2}$   & $-12.1$ &9.944(2)$\times$$10^{-2}$ &$-9.9$ \\
                        ILC    & 1.035(2)$\times$$10^{-2}$  &  9.182(1)$\times$$10^{-3}$  & $-11.3$ & 9.321(4)$\times$$10^{-3}$ & $-9.9$ \\
                        \hline
                \end{tabular}
                \caption{The cross section of $e^{+}e^{-}\rightarrow \mu^+\mu^-H$ with various conditions of the beam polarization and without experimental cuts.}
            \label{tab:4.13}
        \end{center}
\end{table}
\subsection{$e^{+}e^{-}\rightarrow e^+e^-H$}\label{ee}

Note that the notations and symbols used for the $e^{+}e^{-}\rightarrow e^+e^-H$ and other processes are the same as those for the $e^{+}e^{-}\rightarrow \mu^+\mu^-H$ process, for the UP, RR, LL, RL, LR, and ILC.
In this process, because of sizable t-channel amplitude contribution, $\sigma_{\mathcal{O}_\alpha (ILC) e^{+}e^{-}H} = 9.841(2) \times 10^{-3}$ pb is larger than $\sigma_{\mathcal{O}_\alpha (ILC) \mu^{+}\mu^{-}H} = 9.182(1) \times 10^{-3} $ pb in Section $6.1$  approximately $7\%$ at the proposed ILC with  $p_e = - 0.8, p_p = 0.3$. Because it is not needed to form spin-1 state at t-channel, LL and RR cross sections are sizable even at the tree level. Table \ref{tab:4.14} shows detailed values.

\begin{table}[h]
        \begin{center}
                \begin{tabular}{|c||r|r|r|r|r|r|}
                        \hline
                        &      $\sigma_{Tree}$ (pb)    & $\sigma_{\mathcal{O}_\alpha}$ (pb)   &  $\delta_{Total} \%$      & $\sigma_{ISR }$ (pb)    & $\delta_{ISR} \%$ \\
                        \hline
                        UP     & 7.714(2)$\times$$10^{-3}$  &7.348(1)$\times$$10^{-3} $   & $-4.7$ &   6.314(4)$\times$$10^{-3} $&  $-18.1$  \\
                        RR     & 6.87(4)$\times$$10^{-4}  $ & 6.22(1)$\times$$10^{-4} $   & $-9.4$  &   3.080(2)$\times$$10^{-4} $ &$-88.8$ \\
                        LL     & 6.87(4)$\times$$10^{-4}  $   & 6.22(1)$\times$$10^{-4}  $   & $-9.4$  &  3.080(2)$\times$$10^{-4} $ & $-55.4$ \\
                        RL     & 1.149(2)$\times$$10^{-2}$  & 1.236(1)$\times$$10^{-2}$  &7.6        &  9.769(2)$\times$$10^{-3}  $&$-15.0$       \\
                        LR     & 1.799(3)$\times$$10^{-2}$  & 1.576(1)$\times$$10^{-2}$  & $-12.4$  &  1.487(3)$\times$$10^{-2}$ &  $-17.3$ \\
                        ILC    & 1.119(1)$\times$$10^{-2}$  &  9.841(2)$\times$$10^{-3}$& $-12.0$  & 9.159(4)$\times$$10^{-3}$& $-18.1$  \\
                        \hline
                \end{tabular}
                \caption{The cross section of $e^{+}e^{-}\rightarrow e^+e^-H$  with various conditions of the beam polarization and without experimental cuts.}
                 \label{tab:4.14}
        \end{center}
\end{table}

\subsection{$e^{+}e^{-}\rightarrow \tau^+\tau^-H$}

There is a question about tau particle which we are interested in: whether the tau mass can be neglected or not.\\

$\sigma_{\mathcal{O}_\alpha (ILC) \tau^{+}\tau^{-}H}= 9.145(2) \times 10^{-3}$ pb and $\sigma_{\mathcal{O}_\alpha (ILC) \mu^{+}\mu^{-}H}=  9.2100(1) \times 10^{-3} $ pb in Section $6.1$ are quite similar as we expected because all of final radiation effects like $ln \left( \frac{m_\tau^2}{s} \right)$ are canceled according to the Kinoshita-Lee-Nauenberg theorem  \cite{Kinoshita:1962ur},\cite{,Lee:1964is}. The terms from the initial state radiation of $ln \left( \frac{m_e^2}{s} \right)$ of  $e^{+}e^{-}$ still remain, which have the same effects for both processes of $\tau^+\tau^-H$ and $\mu^+\mu^-H$. More results for the case of other  polarization conditions will be shown in Table \ref{tab:4.15}.

\begin{table}[h]
        \begin{center}
                \begin{tabular}{|c||r|r|r|r|r|r|}
                        \hline
                        &      $\sigma_{Tree}$ (pb)    & $\sigma_{\mathcal{O}_\alpha}$ (pb)   &  $\delta_{Total} \%$      & $\sigma_{ISR }$ (pb)    & $\delta_{ISR} \%$ \\
                        \hline
                        UP     & 7.018(2)$\times$$10^{-3}$   &6.708(5)$\times$$10^{-3}$       &$-4.4$ &  6.294(4)$\times$$10^{-3}$  &$-10.3$ \\
                        RR     & 2.10(2)$\times$$10^{-9}$    &  8.44(3)$\times$$10^{-7}$      & 4.0$\times$$10^{4}$  & 8.20(7)$\times$$10^{-12}$    &  $-99.6$\\
                        LL     & 2.10(2)$\times$$10^{-9}$    &  8.44(3)$\times$$10^{-7}$     & 4.0$\times$$10^{4}$  & 8.3(7)$\times$$10^{-12}$     &  $-99.6$\\
                        RL     & 1.1078(2)$\times$$10^{-2}$  &       1.189(6)$\times$$10^{-2}$   &7.3 &  9.17(2)$\times$$10^{-3}$  &$-10.5$        \\
                        LR     & 1.7021(4)$\times$$10^{-2}$  &   1.491(1)$\times$$10^{-2}$  &$-12.4$ & 1.524(2)E$\times$$10^{-2}$   &  $-10.5$ \\
                        ILC    & 1.0343(2)$\times$$10^{-2}$  &        9.145(2)$\times$$10^{-3}$ &$-11.6$ &9.261(4)$\times$$10^{-3}$  & $-10.5$\\
                        \hline
                \end{tabular}
                \caption{The cross section of $e^{+}e^{-}\rightarrow \tau^+\tau^-H$  with various conditions of the beam polarization and without experimental cuts.}
                \label{tab:4.15}
        \end{center}
\end{table}

\subsection{$e^{+}e^{-}\rightarrow  \nu_{\mu} \bar{\nu}_{\mu} H$}

The difference between $\delta_{Total}$ and $   \delta_{ISR}$ is approximately $5\%$. We skip $e^{+}e^{-}\rightarrow  \nu_{\tau} \bar{\nu}_{\tau} H$ process because it is quite similar to this one. More results are provided in  Table \ref{tab:4.16}.
\begin{table}[h]
        \begin{center}
                \begin{tabular}{|c||r|r|r|r|r|r|}
                        \hline
                        &      $\sigma_{Tree}$ (pb)    & $\sigma_{\mathcal{O}_\alpha}$ (pb)   &  $\delta_{Total} \%$      & $\sigma_{ISR }$ (pb)    & $\delta_{ISR} \%$ \\
                        \hline
                        UP     & 1.3891(6)$\times$$10^{-2}$  & 1.328(3)$\times$$10^{-2}$   &$-4.4$ & 1.249(1)$\times$$10^{-2}$  &$-10.1$      \\
                        RR     & 2.71(1)$\times$$10^{-12}$  &  1.669(7)$\times$$10^{-6}$  & 6.16$\times$$10^{7}$  & 1.971(6)$\times$$10^{-12}$  &  $-27.3$ \\
                        LL     & 2.71(1))$\times$$10^{-12}$    &  1.669(7)$\times$$10^{-6}$  & 6.15$\times$$10^{7}$  & 2.307(7)$\times$$10^{-12}$  &      $-14.9$ \\
                        RL     & 2.191(1)$\times$$10^{-2}$   & 2.409(7)$\times$$10^{-2}$ & 10.0 &1.965E(1)$\times$$10^{-2}$  &$-10.3$        \\
                        LR     & 3.367(1)$\times$$10^{-2}$   &  2.905(1)$\times$$10^{-2}$  &  13.7 &   3.022(1)$\times$$10^{-2}$ & $-10.2$\\
                        ILC    & 2.045(1)$\times$$10^{-2}$   & 1.782(1)$\times$$10^{-2}$  & $-12.9$ & 1.837(1)$\times$$10^{-2}$  &        $-10.2$\\
                        \hline
                \end{tabular}
                \caption{The cross section of $e^{+}e^{-}\rightarrow   \nu_{\mu} \bar{\nu}_{\mu} H $  with various conditions of the beam polarization and without experimental cuts.}
                \label{tab:4.16}
        \end{center}
\end{table}

\subsection{$e^{+}e^{-}\rightarrow \nu_{e} \bar{\nu}_{e} H$}

$e^{+}e^{-}\rightarrow \nu_{e} \bar{\nu}_{e} H$ has t-channel. Table \ref{tab:4.17} shows the results of $e^{+}e^{-}\rightarrow \nu_{e} \bar{\nu}_{e} H$ process.
\begin{table}[h]
        \begin{center}
                \begin{tabular}{|c||r|r|r|r|r|r|}
                        \hline
                        &      $\sigma_{Tree}$ (pb)    & $\sigma_{\mathcal{O}_\alpha}$ (pb)   &  $\delta_{Total} \%$      & $\sigma_{ISR }$ (pb)    & $\delta_{ISR} \%$ \\
                        \hline
                        UP     & 2.067(1)$\times$$10^{-2}$  & 2.002(9)$\times$$10^{-2}$  &        $-3.1$ &  1.764(2)$\times$$10^{-2}$ & $-14.6$\\
                        RR     & 3.05(2)$\times$$10^{-12}$   &  3.53(2)$\times$$10^{-6}$  &    1.16$\times$$10^{8}$   & 2.043(8)$\times$$10^{-12}$ & $-33.0$\\
                        LL     & 3.05(2)$\times$$10^{-12}$   &  3.53(2)$\times$$10^{-6}$  &    1.16$\times$$10^{8}$   & 2.385(9)$\times$$10^{-12}$ &$-21.8$ \\
                        RL     & 2.191(1))$\times$$10^{-2}$  &2.414(7)$\times$$10^{-2}$&  10.2   &1.905(1)$\times$$10^{-2}$ &     $-10.3$ \\
                        LR     & 6.076(1))$\times$$10^{-2}$   & 5.612$\times$$10^{-2}$ &$-7.6$    & 5.085(2)$\times$$10^{-2}$&     $-16.3$ \\
                        ILC    & 3.632(1))$\times$$10^{-2}$  & 3.357(3)$\times$$10^{-2}$ & $-7.6$   & 3.044(2)$\times$$10^{-2}$ &$-16.2$  \\
                        \hline
                \end{tabular}
                \caption{The cross section of $e^{+}e^{-}\rightarrow  \nu_{e} \bar{\nu}_{e} H $  with various conditions of the beam polarization and without experimental cuts.} \label{tab:4.17}
        \end{center}
\end{table}

\section{Results of the quark processes}

\subsection{$e^{+}e^{-}\rightarrow  u \bar{u} H$}

QCD corrections up to 4-loop level is well known, and with just in one-loop corrections is popular as the simple factor of $1 + \frac{\alpha_s}{\pi} \simeq 1.03\%$, where $\alpha_s$ is the coupling constant of the strong interaction $(\alpha_s \simeq 1/10)$ thus we will not discuss about it in this paper. Table \ref{tab:4.18} shows calculation results of the first quark process $e^{+}e^{-}\rightarrow  u \bar{u} H$.
\begin{table}[h]
        \begin{center}
                \begin{tabular}{|c||r|r|r|r|r|r|}
                        \hline
                        &      $\sigma_{Tree}$ (pb)    & $\sigma_{\mathcal{O}_\alpha}$ (pb)   &  $\delta_{Total} \%$      & $\sigma_{ISR }$ (pb)    & $\delta_{ISR} \%$ \\
                        \hline
                        UP     & 2.424(1)$\times$$10^{-2}$    & 2.258(2)$\times$$10^{-2}$    & $-6.8$ & 2.178(2)$\times$$10^{-2}$   &     $-10.1$\\
                        RR     & 3.65(2)$\times$$10^{-11}$     &  2.915(1)$\times$$10^{-6}$    & 7.98$\times$$10^{6}$    & 8(4)$\times$$10^{-11}$     &     11.9 \\
                        LL     & 3.65(2)$\times$$10^{-11}$    &  2.92(1)$\times$$10^{-6}$  & 7.98$\times$$10^{6}$    & 8(4)$\times$$10^{-11}$      &     11.9 \\
                        RL     & 3.824(1)$\times$$10^{-2}$    &  4.32(1)$\times$$10^{-2}$    & 13.1 &3.431(1)$\times$$10^{-2}$    &$-10.0$        \\
                        LR     & 5.875(1)$\times$$10^{-2}$    &   4.809(1)$\times$$10^{-2}$   & $-18.2$ &  5.274(1)$\times$$10^{-2}$  &   $-10.0$\\
                        ILC    & 3.570(1)$\times$$10^{-2}$   &    2.934(1)$\times$$10^{-2}$   &  $-17.8$ &3.206(2)$\times$$10^{-2}$    & $-10.0$\\
                        \hline
                \end{tabular}
                \caption{The cross section of $e^{+}e^{-}\rightarrow u \bar{u} H $ with various conditions of the beam polarization and without experimental cuts.}
        \label{tab:4.18}
        \end{center}
\end{table}

\subsection{$e^{+}e^{-}\rightarrow  d \bar{d} H$}

The iso-spin of $u$ quark is up, on the other hand, the iso-spin of $d$ quark is down, which explains the difference between $\sigma_{\mathcal{O}_\alpha (ILC) d \bar{d}H}= 4.070(6)\times 10^{-2} $ pb and $\sigma_{\mathcal{O}_\alpha (ILC )u \bar{u}H}= 2.948(4) \times 10^{-2}$ pb as in the section $7.1$. Table \ref{tab:4.19} shows the results of $e^{+}e^{-}\rightarrow  d \bar{d} H$ process.
\begin{table}[h]
        \begin{center}
                \begin{tabular}{|c||r|r|r|r|r|r|}
                        \hline
                        &      $\sigma_{Tree}$ (pb)    & $\sigma_{\mathcal{O}_\alpha}$ (pb)   &  $\delta_{Total} \%$      & $\sigma_{ISR }$ (pb)    & $\delta_{ISR} \%$ \\
                        \hline
                        UP     & 3.112(1)$\times$$10^{-2}$ & 2.961(1)$\times$$10^{-2}$    & $-4.8$ &    2.796$\times$$10^{-2}$   &$-10.2$  \\
                        RR     & 1.40(1)$\times$$10^{-11}$   & 3.74(2)$\times$$10^{-6}$    & 2.7$\times$$10^{-7}$   &  1.9(8)$\times$$10^{-10}$         &     1257  \\
                        LL     & 1.40(1)$\times$$10^{-11}$  & 3.74(2)$\times$$10^{-6}$    & 2.7$\times$$10^{-7}$   &  1.8(8)$\times$$10^{-10}$     &         1186 \\
                        RL     & 4.907(1)$\times$$10^{-2}$  & 5.205(1)$\times$$10^{-2}$   &6.1       &   6.769(2)$\times$$10^{-2}$  &   $ -10.6$ \\
                        LR     & 7.567(1)$\times$$10^{-2}$  &        6.650(1)$\times$$10^{-2}$& $-12.1$ &  4.402(2)$\times$$10^{-2}$ &  $-10.3$ \\
                        ILC    & 4.581(1)$\times$$10^{-2}$  & 4.070(6)E$\times$$10^{-2}$  &  $-11.5$ &   4.115(2)$\times$$10^{-2}$  &       $-10.2$ \\
                        \hline
                \end{tabular}
                        \caption{The cross section of $e^{+}e^{-}\rightarrow d \bar{d} H $  with various conditions of the beam polarization and without experimental cuts.}
                         \label{tab:4.19}
        \end{center}
\end{table}

\subsection{$e^{+}e^{-}\rightarrow  c \bar{c} H$}

Table \ref{tab:4.20} shows the cross sections of $e^{+}e^{-}\rightarrow  c \bar{c} H$. The mass of charm quark is 1.5 GeV and the mass of up quark is 58 MeV. Because there is no large mass effect from the final state radiation, the results are quite similar as expected.
Because of this observation, we skip $e^{+}e^{-}\rightarrow  s \bar{s} H$ process because this process has the same characteristic of $e^{+}e^{-}\rightarrow  d \bar{d} H$ process.

\begin{table}[h]
        \begin{center}
                \begin{tabular}{|c||r|r|r|r|r|r|}
                        \hline
                        &      $\sigma_{Tree}$ (pb)    & $\sigma_{\mathcal{O}_\alpha}$ (pb)   &  $\delta_{Total} \%$      & $\sigma_{ISR }$ (pb)    & $\delta_{ISR} \%$ \\
                        \hline
                        UP     & 2.421(1)$\times$$10^{-2}$  &  2.259(1)$\times$$10^{-2}$&  $-6.8$  & 2.175(2)$\times$$10^{-2}$  &$-10.2$ \\
                        RR     & 1.482(7)$\times$$10^{-11}$  &  2.91(1)$\times$$10^{-6}$  & 1.96$\times$$10^{7}$&  1.4155(8)$\times$$10^{-11}$ &        $-4.9$ \\
                        LL     & 1.482(7)$\times$$10^{-11}$  &  2.91(1)$\times$$10^{-6}$   & 1.96$\times$$10^{7}$    & 1.4600(8)$\times$$10^{-11}$  &        $-1.5$ \\
                        RL     & 3.819(1)$\times$$10^{-2}$  &       4.300(1)$\times$$10^{-2}$&  12.6  &  3.425(2)$\times$$10^{-2}$& $-10.3$   \\
                        LR     & 5.890(1)$\times$$10^{-2}$  &  4.771(1)$\times$$10^{-2}$  &   $-19.0$ & 5.266(1)$\times$$10^{-2}$   &        $-10.6$ \\
                        ILC    & 3.566(1)$\times$$10^{-2}$  &  2.926(4)$\times$$10^{-2}$ & $-18.0$ &  3.202(2)$\times$$10^{-2}$ &  $-10.2$ \\
                        \hline
                \end{tabular}
                        \caption{The cross section of $e^{+}e^{-}\rightarrow c \bar{c} H $  with various conditions of the beam polarization and without experimental cuts.}
                 \label{tab:4.20}
        \end{center}
\end{table}

\subsection{$e^{+}e^{-}\rightarrow  b \bar{b} H$}

Table \ref{tab:4.21} summaries cross sections of $e^{+}e^{-}\rightarrow  b \bar{b} H$. We keep the bottom-Yukawa coupling with NOLFS approximation for $e^{+}e^{-}\rightarrow  b \bar{b} H$ but not in other processes. Let's compare the results of $e^{+}e^{-}\rightarrow  d \bar{d} H$ and $e^{+}e^{-}\rightarrow  b \bar{b} H$ of the proposed ILC, $p_e = - 0.8, p_p = 0.3$ at the $\delta_{Total(ILC)} $. which are $-11.5\%$ and $12.5\%$, respectively. This $1\%$ difference is caused by the bottom-Yukawa coupling. With ISR, the results for $e^{+}e^{-}\rightarrow  d \bar{d} H$ and $e^{+}e^{-}\rightarrow  b \bar{b} H$ are quite similar as expected. This difference between  $\delta_{Total(ILC)}$  and $\delta_{ISR(ILC) }$ is very interesting. Because this $1\%$ difference originated from the bottom-Yukawa coupling of the bottom quark, though it maybe hard to observe at the experiments.

\begin{table}[h]
        \begin{center}
                \begin{tabular}{|c||r|r|r|r|r|r|}
                        \hline
                        &      $\sigma_{Tree}$ (pb)    & $\sigma_{\mathcal{O}_\alpha}$ (pb)   &  $\delta_{Total} \%$      & $\sigma_{ISR }$ (pb)    & $\delta_{ISR} \%$ \\
                        \hline
                        UP     & 3.078(1)$\times$$10^{-2}$   &  2.88(1)$\times$$10^{-2}$ & $-6.3$ &  2.765(2)$\times$$10^{-2}$ &$-10.2$    \\
                        RR     & 7.40(4)$\times$$10^{-12}$   &  3.69(2)$\times$$10^{-6}$  &  5.0$\times$$10^{7}$  &  5.70(3)$\times$$10^{-12}$  & $-22.9$  \\
                        LL     & 7.40(4)$\times$$10^{-12}$    &  3.69(2)$\times$$10^{-6}$   &  5.0$\times$$10^{7}$  &  6.31(3)$\times$$10^{-12}$  & $-14.6$       \\
                        RL     & 4.855(1)$\times$$10^{-2}$   & 5.06(1)$\times$$10^{-2}$ & 4.2  & 4.035(1)$\times$$10^{-2}$  &      $-10.2$\\
                        LR     & 7.461(2)$\times$$10^{-2}$   & 6.46(1)$\times$$10^{-2}$  & $-13.5$ &  6.691(2$\times$$10^{-2}$ & $-10.2$ \\
                        ILC    & 4.533(1)$\times$$10^{-2}$   & 3.969(4)$\times$$10^{-2}$  & $-12.5$  &  4.068(2)E$\times$$10^{-2}$ & $-10.3$ \\
                        \hline
                \end{tabular}
                \caption{The cross section of $e^{+}e^{-}\rightarrow b \bar{b} H $  with various conditions of the beam polarization and without experimental cuts.}
                \label{tab:4.21}
        \end{center}
\end{table}

\noindent
\section{Recoil mass distribution of $e^{+}e^{-}\rightarrow \mu^+\mu^-H$ with beam polarization effects}

In this section, we discuss the ZH recoil mass distribution $ZH \rightarrow \mu^+\mu^-H$ with $\mathcal{O}(\alpha)$ corrections and beam polarization effects. To conduct a more realistic analysis, we apply the following three experimental cuts:\\
\noindent
1. The angular cuts on $ \theta_{\mu^+ }, \theta_{\mu^-}$
\begin{equation}
        10^o < \theta_{\mu^+ }, \theta_{\mu^-} <170^o,
\end{equation}
where $\theta_{\mu^+ } $ and $ \theta_{\mu^-}$ are the scattering angles of anti-muon and muon, respectively. \\
2. The energy cuts on   $E_{\mu^+}, E_{\mu^-} $
\begin{equation}
        E_{\mu^+}, E_{\mu^-} >10 ~ \text{GeV},
\end{equation}
where   $E_{\mu^+}$ and $ E_{\mu^-} $ are the energies of anti-muon and muon, respectively.\\
\noindent
3. The invariant mass cut $M_{\mu^+ \mu^-}$
\begin{equation}
        m_Z-3\Gamma_Z < M_{\mu^+ \mu^-} < m_Z+3\Gamma_Z.
\end{equation}
where $M_{\mu^+ \mu^-}$ is the invariant mass of anti-muon and muon particles; $\Gamma_Z = 2.49$ GeV  \cite{PhysRevD.98.030001} is the width of the Z-boson.\\

\begin{figure}[h!]
       \begin{center}
                \includegraphics[height =2.5in, angle =0]{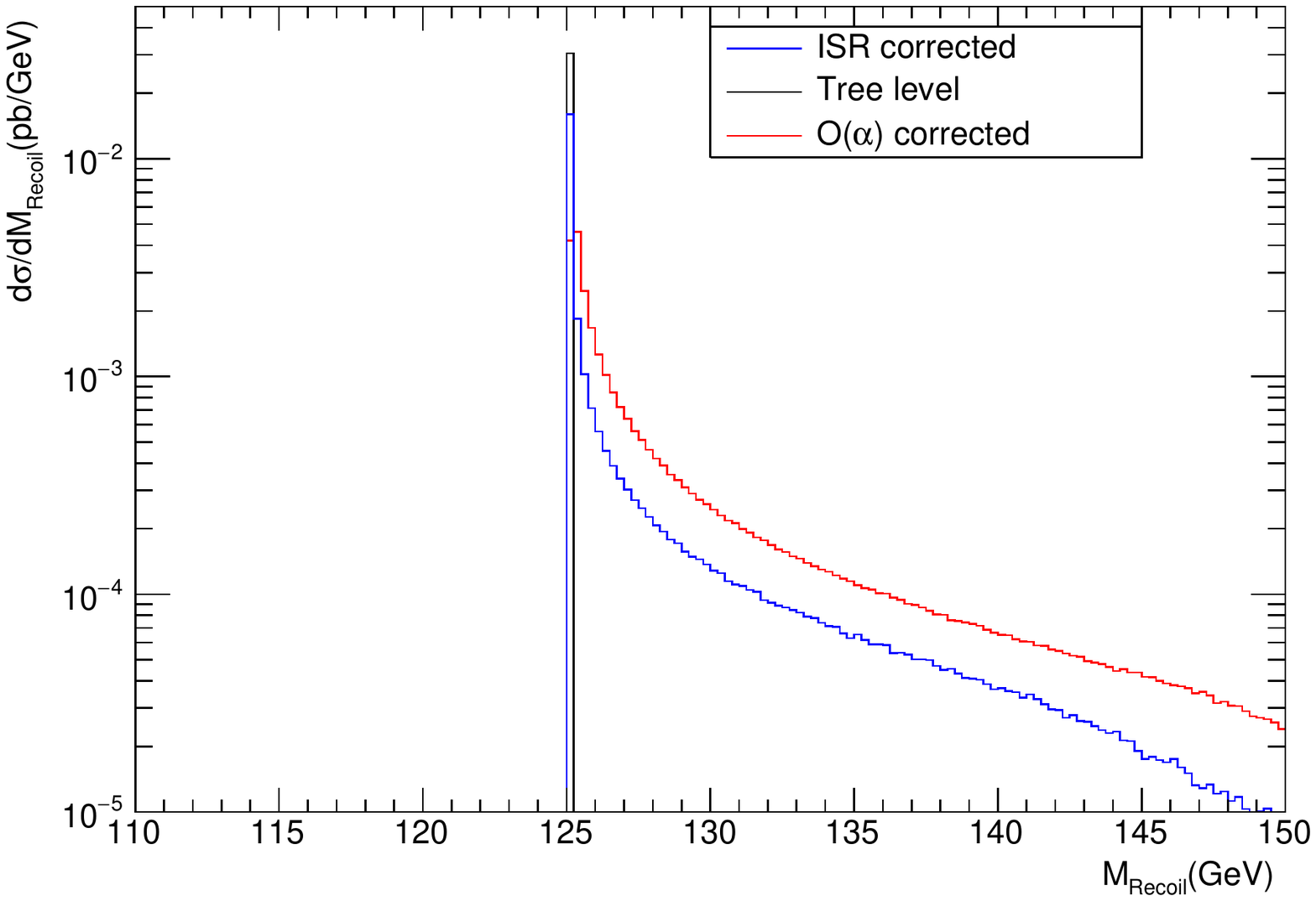}
                \caption{Recoil mass distribution of $e^{+}e^{-}\rightarrow \mu^+\mu^-H$  after applying the experimental cuts as in Eqs (4.3), (4.4),(4.5) at $\sqrt{_s}=250$ GeV. The bin width is 0.3 GeV.}
                \label{fig.5}
        \end{center}
\end{figure}

Next, we present the results of the ZH recoil mass distribution with the $\mathcal{O}(\alpha)$ corrections of  $e^{+}e^{-}\rightarrow \mu^+\mu^-H$ at $\sqrt{_s}=250$ GeV, after applying the three types of experimental cuts as defined above. The recoil mass distribution is shown in Fig. \ref{fig.5}.\\

\noindent
The black line shows the tree level distribution, the blue line shows the ISR corrected distribution and the red line shows the $\mathcal{O}(\alpha)$ corrected distribution. The black line only shows the one bin peak and of which position
indicates the mass of Higgs boson. On including the $\mathcal{O}(\alpha)$ corrections,
a tail structure appears owing to the ISR effects, which is called
the radiative tail. The height of peak is substantially reduced owing to the one-photon emission effect. However, the ISR effects are included, the height of the peak increases again because of the higher-order radiation effect.
This swing back effect can be seen, occasionally when comparing the
higher-order and $\mathcal{O}(\alpha)$ corrections.\\

After applying the three experimental cuts, we obtained the total cross sections at $\mathcal{O}(\alpha)$ and those including ISR effects and the total ratios at $\mathcal{O}(\alpha)$ and those including ISR effects as shown in Table \ref{tab:4.12}

\begin{table}[h]
        \begin{center}
                \begin{tabular}{|c|r|r|r|r|r|}
                        \hline
                          $\sigma_{Tree}$ (pb)  &  $\sigma_{\mathcal{O}_\alpha}  $ (pb) & $\sigma_{ISR}  $ (pb) &  $\delta_{Total}  \%$ & $\delta_{ISR}  \%$\\
                        \hline
                            9.143(1)$\times 10^{-3}$    &  7.910(1)$\times 10^{-3}$  & 8.236(4)$\times 10^{-3}$ & $-13.2$ & $-9.9$   \\
                        \hline
                \end{tabular}
                \caption{Summary table of cross sections and total ratios of $e^{+}e^{-}\rightarrow \mu^+\mu^-H$ with experimental cuts and with ILC proposed polarization with $p_e = -0.8$ , $p_p = 0.3$.}
                \label{tab:4.12}
        \end{center}
\end{table}
\noindent
\section{Discussion} 

First, we discuss the behavior of high-order corrections for the $e^{+}e^{-}\rightarrow \mu^+\mu^-H$ process in detail. For this process, various conditions of beam polarization without experimental cuts are shown in Table \ref{tab:4.13}.
At the tree level, the cross sections corresponding to the RR and LL are negligibly small compared with those of RL and LR since RR and LL polarizations cannot produce spin-1 boson. 
Small cross sections of RR and LL polarizations are resulted from spin-flip effect, which are propotional to  $|\frac{me^2}{s}| \propto \left(\frac{10^{-3}}{10^2} \right)^2 \sim 10^{-10}$. These results are consistent with $\frac{LL+RR}{LR+RL} \sim 10^{-11} $.
	Moreover,  $\sigma_{\mathcal{O}_\alpha} $ gives sizeable cross sections on RR and LL polarizations, because $\mathcal{O}_\alpha $ amplitude includes spin-flip diagrams of the initial state radiation.
	For LL and RR, although the cross sections of LL and RR at the tree level are very small, the very large total ratios  $\delta_{Total} \sim 10^{6}$ are observed. 

	 $\sigma_{\mathcal{O}_\alpha(ILC)\mu^+\mu^- H} = 9.182(1) \times 10^{-2}$ pb is larger than  $\sigma_{\mathcal{O}_\alpha(UP)\mu^+\mu^- H} = 6.724(1) \times 10^{-3}$ pb thanks to the advantages of the beam polarization of the linear collider. It has significantly better statistics of Higgs with the same luminosity. \\
	 
Next, we discuss the effect of higher-order corrections on weak interactions and photonic interactions. Note that
     $\sigma_{\mathcal{O}_\alpha}$ includes both QED and weak corrections. However, it is impossible in general to clearly separate theses two effects. we calculated $\delta_{ISR}$ as an approximation for QED corrections because ISR corrections dominate over all QED corrections. In this approximation, the weak corrections can be estimated as 
     $ \sigma_{weak} = \sigma_{\mathcal{O}_\alpha} - \sigma_{ISR}$.
    $\delta_{ISR(RL)}$ and $\delta_{ISR(LR)}$  were obtained to be $-19.3\%$ and $-9.9\%$, respectively; however, $\delta_{Total(RL)} = 7.7\%$, and $\delta_{Total(LR)} = -12.1\%$. Thus the pure weak corrections corresponding to (RL) is $27\%$ but that corresponding to (LR) is only $-2 \%$. From these results, it is clear that the experiements with  beam polarization needs the  precise calculations of the electroweak $\sigma_{\mathcal{O}_\alpha}$ corrections.
	$\delta_{Total(ILC)} = -11.3\%$, $\delta_{ISR(ILC)} = -9.9\%$, and the weak correction for the proposed ILC polarization, $p_e = - 0.8, p_p = 0.3$ is approximately $-1\%$. This result is sufficient to motivate the ILC to consider the contribution of the pure weak correction. Additionally $\delta_{weak}$ are necessary to compare with the naive estimation from the running alpha effects of $+14 \%=(+7)\%+(+7)\%$.\\

\begin{table}[h]
	\begin{center}
		\begin{tabular}{|c||r|r|r|r|r|r|}
			\hline
			Processes	&      $\sigma_{Tree}$ (pb)    & $\sigma_{\mathcal{O}_\alpha}$ (pb)   &  $\delta_{Total} \%$      & $\sigma_{ISR }$ (pb)    & $\delta_{ISR} \%$ \\
			\hline
			$\mu^+\mu^-H$                   & 1.035(2)$\times10^{-2}$  & 9.182(1)$\times10^{-3}$   &$-11.3$& 9.321(4)$\times10^{-3}$  &$-9.9$ \\
			$e^+e^-H$                       & 1.119(1)$\times10^{-2}$  & 9.841(2)$\times10^{-3}$  & $-12.0$ &	9.159(4)$\times10^{-3}$  &$-18.1$ \\
			$\tau^+\tau^-H$                 & 1.034(1)$\times10^{-2}$  & 9.145(2)$\times10^{-3}$	 & $-11.6$ &	9.261(4)$\times10^{-3}$ &$-10.5$ \\
			$\nu_{\mu} \bar{\nu}_{\mu} H$   & 2.045(1)$\times 10^{-2}$  &  1.782(1)$\times10^{-2}$  & $-12.9$ &	1.837(1)$\times10^{-2}$  &$-10.2$ \\
			$\nu_{e} \bar{\nu}_{e} H$       & 3.632(1)$\times10^{-2}$  & 3.357(3)$\times10^{-2}$   & $-7.6$ &	3.044(2)$\times10^{-2}$  &$-16.2$ \\
			\hline
		\end{tabular}
		\caption{Lepton processes with the ILC proposed beam polarization with $p_e = - 0.8, p_p = 0.3$ at $\sqrt{s} = 250$ GeV and without experimental cuts.}
			\label{tab:5.2}
	\end{center}
\end{table}

We have summarized lepton processes $e^{+}e^{-}\rightarrow f\bar{f}H$ (i.e., $\mu^+\mu^-H$, $e^+e^-H$, $\tau^+\tau^-H$, $\nu_{\mu} \bar{\nu}_{\mu} H$, and $\nu_{e} \bar{\nu}_{e} H$)   in Table  \ref{tab:5.2} .

\begin{figure}[h!]
	\begin{center}
		\includegraphics[height =3.5in, angle =0]{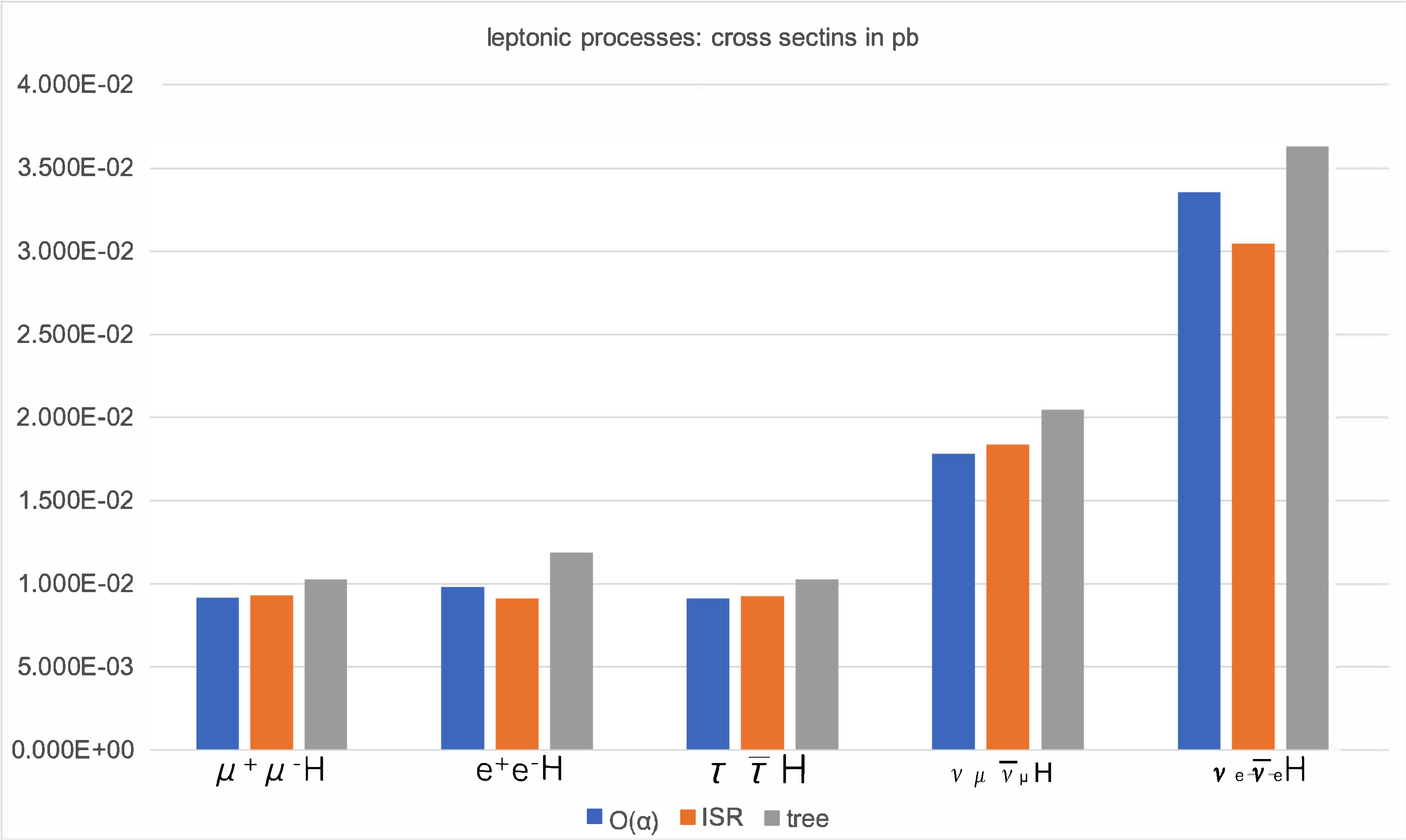}
		\caption{Cross sections of the leptonic processes with the ILC proposed beam polarization, $p_e = - 0.8, p_p = 0.3$ at $\sqrt{s} = 250$ GeV and without experimental cuts.}
		\label{fig.5.1}
	\end{center}
\end{figure}

\begin{figure}[h!]
	\begin{center}
		\includegraphics[height =3.5in, angle =0]{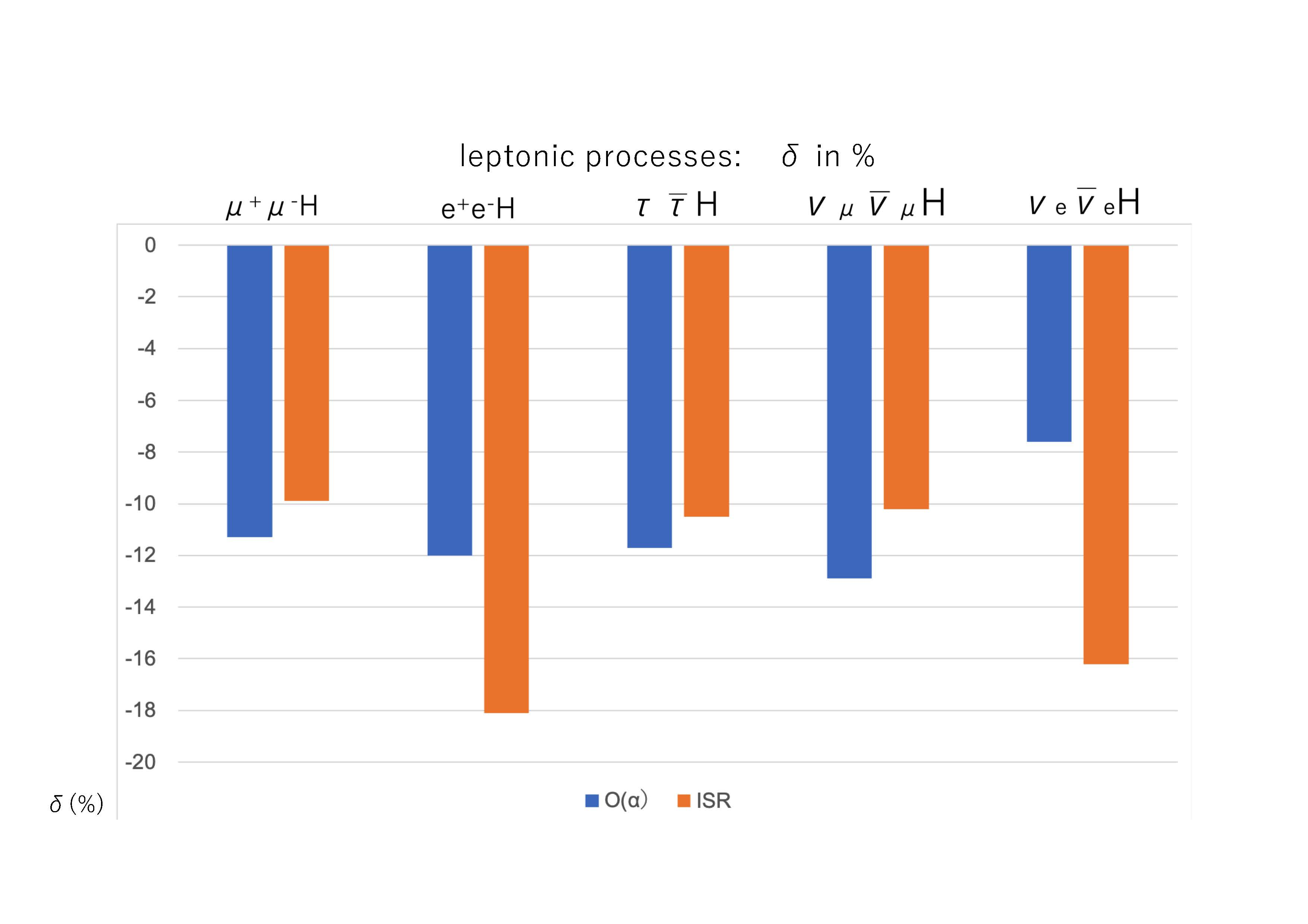}
		\caption{Ratios of the leptonic processes with the ILC proposed beam polarization, $p_e = - 0.8, p_p = 0.3$ at $\sqrt{s} = 250$ GeV and without experimental cuts.}
		\label{fig.5.2}
	\end{center}
\end{figure}

We present two separate figures for the cross sections and total ratios, for convenience. Fig. \ref{fig.5.1} shows the cross sections at the tree level, as well as $\sigma_{\mathcal{O}_\alpha}$, and ISR cross sections. Fig. \ref{fig.5.2} shows $\delta_{Total}$ and $\delta_{ISR}$ of the four leptonic processes.

Owing to the sizable contribution of the t-channel amplitude, $\sigma_{\mathcal{O}_\alpha(ILC) \nu_e\bar\nu_e H} = 3.632(1) \times 10^{-2}$ pb is larger than $\sigma_{\mathcal{O}_\alpha (ILC) \nu\mu \bar\nu_\mu H} = 1.782(1) \times 10^{-2} $ pb. $\delta_{Total(ILC) \nu_\mu \bar\nu_\mu H} = -12.9\%$ and $\delta_{ISR(ILC)\nu_\mu \bar\nu_\mu H} = -10.2\%$, and hence, the pure weak correction for $\nu_\mu \bar\nu_\mu H$ is approximately $-3\%$ whereas, $\delta_{Total(ILC) \nu_e \bar\nu_e H} = -7.6\%$ and $\delta_{ISR(ILC)\nu_e \bar\nu_e H} = -16.2\%$, hence the pure weak corrections for $\nu_e \bar\nu_e H$ is approximately $+9\%$. Moreover, $\mu^+\mu^-H$ and $e^+e^-H$ have the same tendency as 
	 $\nu_{\mu} \bar{\nu_\mu} H$ 	 and $\nu_{e} \bar{\nu}_{e} H$. 
 	 From this discussion, it is clear that although t-channel cross sections at the tree level are small, they are important for

    $\delta_{Total (ILC) \nu_\mu \bar\nu_\mu H} = -12.9\%$ and $\delta_{Total  (ILC)  \nu_e \bar\nu_e H} = -7.6 \%$; this $5.3\%$ difference maybe owing to the charge/neutral channel or iso-spin up/down type. 

Next, we discuss the quark processes. $e^{+}e^{-}\rightarrow f\bar{f}H$ (i.e., $u \bar{u} H$, $d \bar{d} H$, $c \bar{c} H$, $b \bar{b} H$), which are summarized in the Table \ref{tab:5.3}.

\begin{table}[h]
	\begin{center}
		\begin{tabular}{|c||r|r|r|r|r|r|}
			\hline
			Processes	&      $\sigma_{Tree}$ (pb)    & $\sigma_{\mathcal{O}_\alpha}$ (pb)   &  $\delta_{Total} \%$      & $\sigma_{ISR }$ (pb)    & $\delta_{ISR} \%$ \\
			\hline
			$u \bar{u} H$     & 3.570(1)$\times10^{-2}$  & 2.934(1)$\times10^{-2}$   & -17.8  &	3.206(2)$\times10^{-2}$   &	$-10.0$\\
			$d \bar{d} H$     & 4.581(1)$\times10^{-2}$  & 4.070(1)$\times10^{-2}$   & -11.5  &	4.115(2)$\times10^{-2}$   &	$-10.2$\\
			$c \bar{c} H$     & 3.566(1)$\times10^{-2}$  & 2.926(1)$\times10^{-2}$   & -18.0	  &    3.215(2)$\times10^{-2}$   &       $-10.2$\\
			$b \bar{b} H$     & 4.533(1)$\times10^{-2}$  & 3.969(4)$\times10^{-2}$   & -12.5  &	4.068(2)$\times10^{-2}$   &	$-10.3$\\
			\hline
		\end{tabular}
		\caption{Quark processes with the ILC proposed beam polarization, $p_e = - 0.8, p_p = 0.3$ at $\sqrt{s} = 250$ GeV and without experimental cuts.}
		\label{tab:5.3}
	\end{center}
\end{table}

\begin{figure}[h!]
	\begin{center}
		\includegraphics[height =3.5in, angle =0]{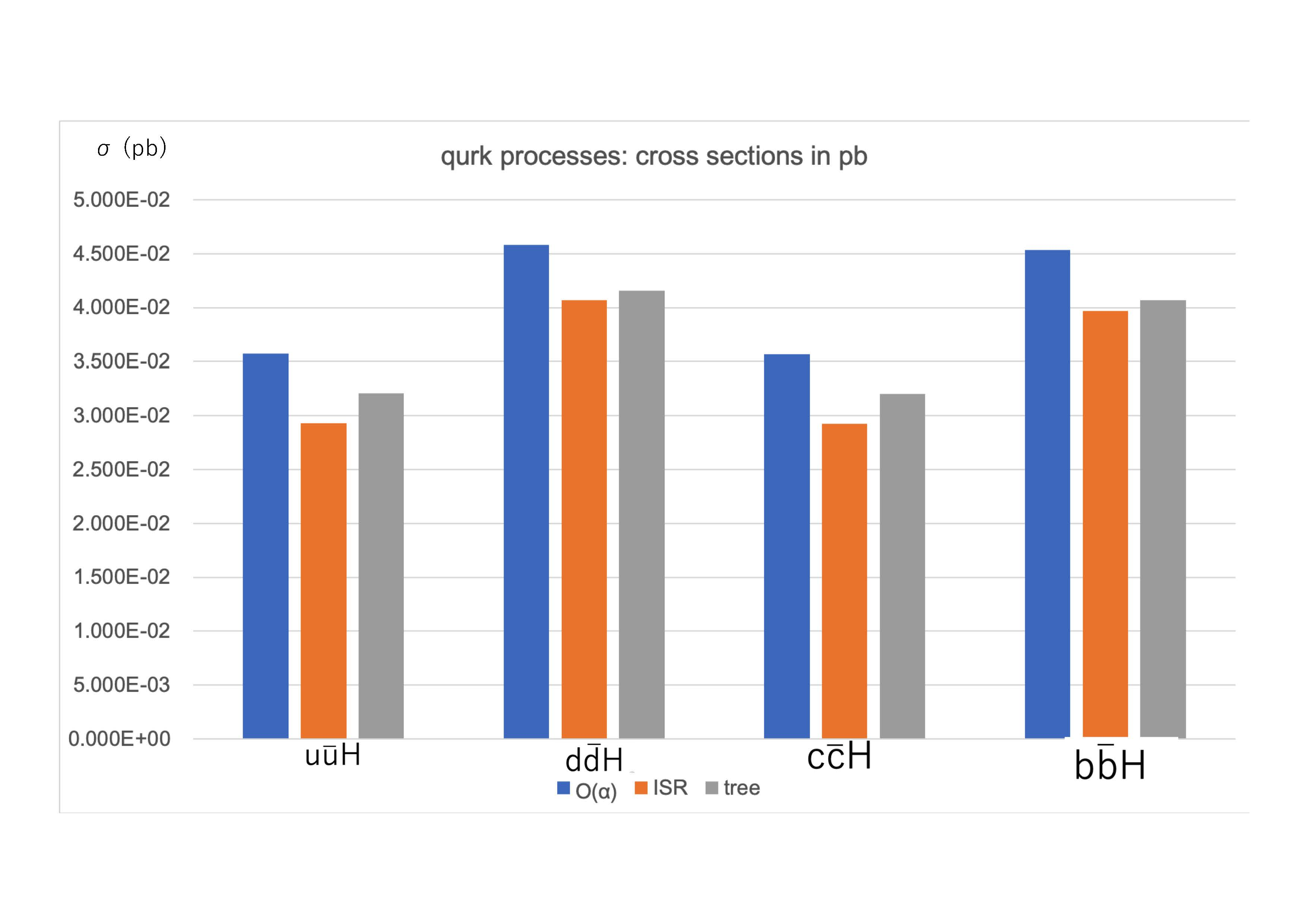}
		\caption{Cross sections of the quark processes with the ILC proposed beam polarization, $p_e = - 0.8, p_p = 0.3$ at $\sqrt{s} = 250$ GeV and without experimental cuts.}
		\label{fig.5.3}
	\end{center}
\end{figure}

\begin{figure}[h!]
	\begin{center}
		\includegraphics[height =3.5in, angle =0]{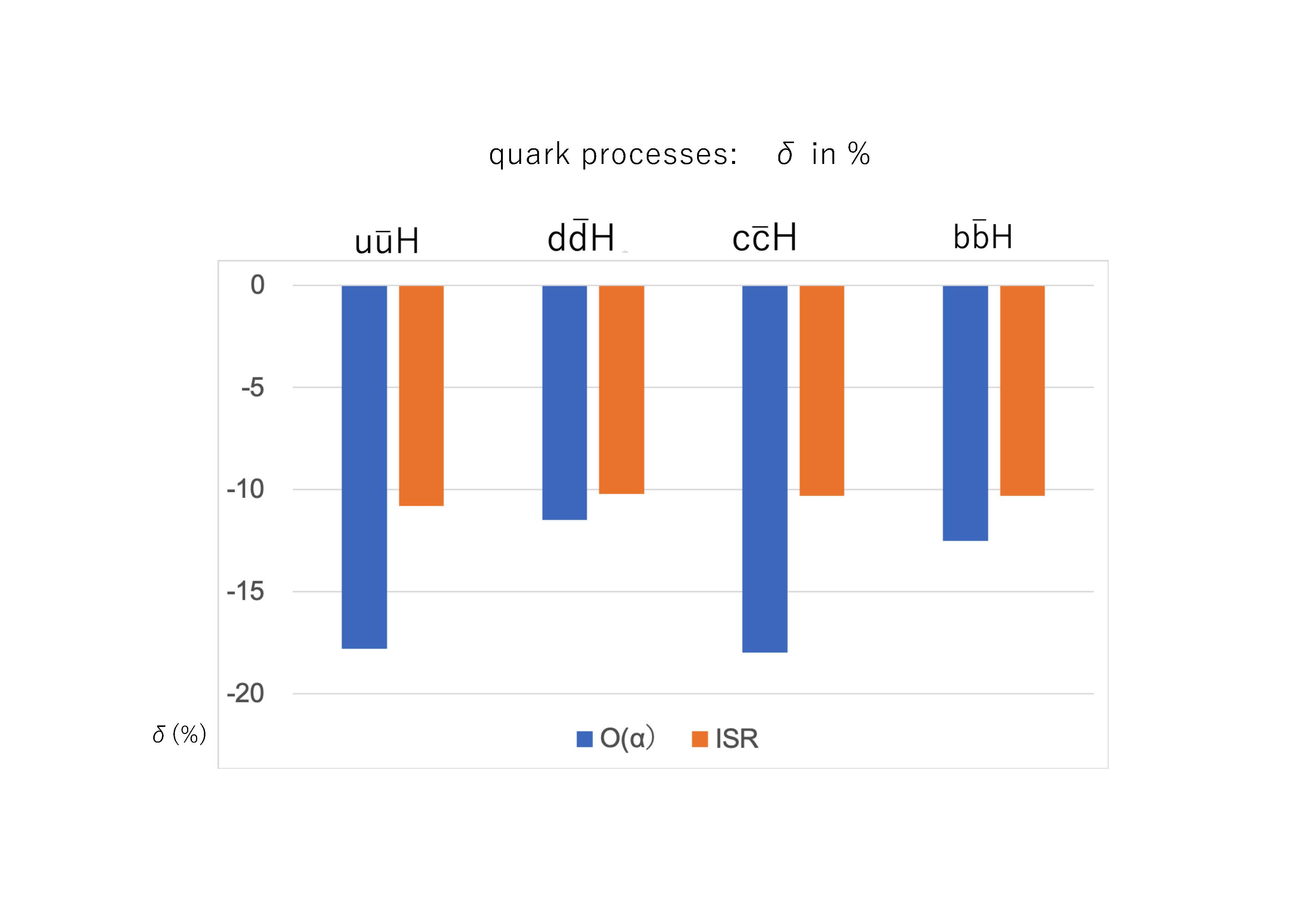}
		\caption{Ratios of the quark processes with the ILC proposed beam polarization $p_e = - 0.8, p_p = 0.3$ at $\sqrt{s} = 250$ GeV and without experimental cuts.}
		\label{fig.5.4}
	\end{center}
\end{figure}

Again, we present two separate figures for the cross sections and total ratios. Fig. \ref{fig.5.3} shows the cross sections at the tree level, as well as the  $\sigma_{\mathcal{O}_\alpha}$, and ISR cross sections. Fig  \ref{fig.5.4} shows $\delta_{Total}$ and $\delta_{ISR}$ of the four quark processes. 

 $\delta_{ISR}$ is approximately $-10\%$ for all processes.
	
	$\delta_{Total}$ of the up-type quarks is $-18\%$, while $\delta_{Total}$ of the down-type quarks is $-12\%$ because of the charge difference or up/down type with different iso-spins.
	
	 $\sigma_{\mathcal{O}_\alpha(ILC)u\bar{u}H} = 2.933(4) \times 10^{-2}$ pb
	is larger than 
	$\sigma_{\mathcal{O}_\alpha(ILC)c\bar{c}H} = 2.926(4)  \times 10^{-2}$ pb
	as expected. These exhibit the opposite tendency to leptons.
	 
	 $\delta_{Total(ILC)d\bar{d}H}= -11.5 \%$ and $\delta_{Total(ILC) b\bar{b}H}= -12.5 $, with the 1$\%$ difference arising  because of the Yukawa coupling. Weak corrections for the up-type and down-type quarks are approximately $-8\%$ and $-2\%$, respectively.  \\

Next, we compare the lepton and quark channels. For leptons, $\delta_{ISR}$ for s-channel and t-channel processes proximate $-10\%$ and $-18\%$, respectively; for quarks,  $\delta_{ISR}$ is approximately $-10\%$ for all the cases with the ILC polarized beams and without experimental cuts.

Comparing the lepton and quark channels, the difference between the up-type and the down-type are clearly observed in quarks; however, this is not observed for lepton.\\

We also calculated the recoil mass distribution after applying three experimental cuts for the $e^{+}e^{-}\rightarrow \mu^+\mu^-H$ process at the $\mathcal{O}(\alpha)$ corrections and including the ISR effects. The following results were obtained: $\delta_{Total} = -13.2 \%$ and $\delta_{ISR} = -9.9 \%$, such that the weak correction was estimated to be $-3\%$. Thus, to measure the $g_{HZZ}$ coupling within 1$\%$ accuracy, the weak correction of $-3\%$ cannot be neglected.\\

\section{Conclusion} 

We presented the full ${{\cal O}}(\alpha)$
electroweak radiative corrections to associated 9 Higgs fermion pair
production in $e^+e^-$ collisions with the beam polarization at the International Linear Collider(ILC).
The computation was performed with the help of {\tt GRACE-loop}. 
We compared the full ${{\cal O}}(\alpha)$ electroweak radiative corrections with the factorized initial state radiation(ISR) effects
to estimate the pure weak corrections. The matrix element to be produced by  {\tt GRACE-loop} was tested by (1)
the renormalization,(2)infrared divergence-free, (3)the hard photon cut parameters ($k_c$) independence in $0.1\%$, (4)
independence of the non-linear gauge parameters. 

Our calculations have the following features: (1) the beam polarization effect was considered, (2) the mass effect of all the fermions except the neutrinos was retained, and (3) Yukawa coupling of the bottom quark was included. To the best of our knowledge, this is the first detailed discussion of the $e^+e^- \rightarrow f\bar{f} H$  processes. Furthermore, we compared our results with those of multiple photon emission from electrons and positrons in the ISR process and highlighted that the $\mathcal{O}(\alpha)$ corrections presented in this work are important for the analysis of experimental data. In the case of the $e^{+}e^{-}\rightarrow \mu^+\mu^-H$ process with the experimental cuts and the proposed ILC beam polarization condition, the pure weak correction was estimated to be $-3\%$. Thus, in order to measure the $g_{HZZ}$ coupling within 1$\%$ accuracy, the pure weak $\mathcal{O}(\alpha)$ corrections cannot be neglected.

\vspace{0.5cm}
 {\bf Acknowledgments} \\
\noi This work is a part of the collaboration in the Minami-Tateya group in the IPNS from KEK.
We acknowledge useful discussions with T. Kaneko and F. Yuasa from KEK and their continuous encouragement.
Junpei Fujimoto and Nhi M. U. Quach especially acknowledge Z. Hioki for his supplying the FOTRAN program, {\tt MW.f}, and the best-recomended values of the light-quark masses to calculate the electroweak $\mathcal{O}(\alpha)$ corrections of the mass of W boson.
Nhi M. U. Quach also would like to acknowledge Tran Thanh Van for his help after getting her Ph.-D.
Yoshimasa Kurihara appreciates the warm hospitality of a member of the Nikhef theory group,
especially J. Vermaseren and E. Laenen. He started the first calculation of
this study during his stay at Nikhef.

\bibliographystyle{ptephy}
\bibliography{2204-028-3C-JunpeiFujimoto}

\end{document}